\documentclass[aps,prd,amsmath,amsfonts,amssymb,groupedaddress,nofootinbib,nobalancelastpage,floatfix,secnumarabic,superscriptaddress,secnumarabic,preprintnumbers]{revtex4-2}
\usepackage[a4paper,left=1.3in,right=1.3in,top=1.2in,bottom=1.2in]{geometry}
    
\usepackage{epsfig, graphicx, hyperref, slashed,xspace}
\usepackage[dvipsnames]{xcolor}
\usepackage[utf8]{inputenc}
\usepackage[T1]{fontenc}
\usepackage{cleveref}
\usepackage{soul}
\usepackage{diagbox}
\usepackage{booktabs}
\usepackage{multirow}
\usepackage{setspace}
\usepackage{float}
\usepackage{caption}
\usepackage{subcaption}

\hypersetup{colorlinks=true,linkcolor=Maroon,citecolor=ForestGreen,filecolor=ForestGreen,urlcolor=ForestGreen}

\newcommand{\MG}{{\sc MadGraph5\_aMC@NLO}\xspace}
\newcommand{\pythiaN}{{\sc Pythia8}\xspace}
\newcommand{\pythia}{{\sc Pythia6}\xspace}
\newcommand{\whizard}{{\sc Whizard3}\xspace}

\newcommand{\fr}{{\sc FeynRules}\xspace}
\newcommand{\ufo}{{\sc UFO}\xspace}
\newcommand{\ohsm}{\texttt{DM\_OHSM}\xspace}
\newcommand{\ab}{{$\rm{ab}^{-1}$}\xspace}
\newcommand{\GeV}{{\ensuremath\,\rm GeV}\xspace}

\makeatletter\def\l@subsubsection#1#2{}\makeatother
\usepackage{orcidlink}

\begin{document}

\preprint{RBI-ThPhys-2026-19}

\title{Search for Invisibly Decaying Light Scalars at the FCC-ee}

\author{Aman Desai \orcidlink{0000-0003-2631-9696}}\email{aman.desai@adelaide.edu.au}
\affiliation{Department of Physics, Adelaide University, Adelaide, SA 5005, Australia}

\author{Tania Robens \orcidlink{0000-0002-9913-5225}}
\email{tania.natalie.robens@irb.hr}
\affiliation{Ruder Boskovic Institute, Bijenicka cesta 54, 10000 Zagreb, Croatia}

\date{\today}

\begin{abstract}

\vspace{3mm}
\centerline{\bf \large Abstract}
\vspace{3mm}
We investigate the production of invisibly decaying light scalars in association with hadronically decaying $Z$ bosons at the Future Circular Collider-ee at a centre-of-mass energy $\sqrt{s}=240$ GeV. Several new physics models predict the existence of these low-mass scalar states, while the existing experimental constraints do not yet exclude these states. We study the low-mass scalar based on a simplified extension of the Standard Model, introducing an additional scalar singlet and a scalar dark matter candidate.  The analysis is performed for a set of new scalars with mass in the range $(15, 120)$ GeV, by employing a selection-based strategy complemented with Multivariate Analysis techniques to discriminate the signal from background. The expected upper limits on the production cross-section times the branching fraction of the new scalars decaying invisibly are evaluated as a function of the scalar mass. We find that sensitivities of $\sim 10^{-2}$--$10^{-1}$~fb are achievable for scalar masses below $\sim 60$ \GeV, while sensitivities of $0.1$--$1$~fb are obtained for masses between $\sim 60$ and $120$ \GeV, with the weakest sensitivity near the $Z$-pole at scalar mass $M_{S}\approx 90$ \GeV. Depending on the mixing angle, novel scalars with masses below 80 \GeV as well as in the range 110--120 \GeV are within discovery reach.

\end{abstract}

\maketitle

\newpage
\section{Introduction}\label{sec:intro}

Since the discovery of the Higgs boson~\cite{ATLAS:2012yve,CMS:2012qbp}, particle physics has entered an interesting era. So far, most current measurements, however, align quite well with Standard Model (SM) predictions~\cite{ParticleDataGroup:2026aaa}. On the other hand, the SM cannot explain several features that are currently observed in nature, e.g., the source of dark matter (DM) or CP violation that leads to the currently observed matter-antimatter asymmetry in the universe. For this, additional fields need to be present that modify the Lagrangian of the SM and introduce new dynamics. One possible extension is the enhancement of the SM scalar sector. In fact, many extensions of the Standard Model in particle physics predict the existence of scalars in addition to the Standard Model Higgs boson (see e.g., \cite{Robens:2025nev} for a recent review).

A decay-mode-independent measurement was also performed by the OPAL experiment at LEP, although its sensitivity was limited by the lower integrated luminosity~\cite{OPAL:2002ifx}. Searches for an additional neutral scalar $S$ produced via $e^+e^-\to ZS$ and decaying invisibly have been studied at CLIC ($\sqrt{s}=380~\GeV$)~\cite{Mekala:2020zys}, the ILC ($\sqrt{s}=250~\GeV$)~\cite{Brudnowski:2026fms}, and CEPC ($\sqrt{s}=250~\GeV$)~\cite{Ai:2025cpj}. The CLIC and ILC studies reconstruct hadronic decays of the associated $Z$ boson, whereas the CEPC study considers leptonic $Z$ decays. For recent broader overviews of searches for additional scalars and the channels considered in the ECFA study prepared as input to the European Strategy, see Refs.~\cite{Robens:2026xfc} and \cite{deBlas:2024bmz,Altmann:2025feg}, respectively\footnote{See also \cite{Drechsel:2018mgd} for projections based on a LEP search recast.}.

One of the top priorities in the current particle landscape after the HL-LHC program is the construction of a proposed Higgs factory with a centre-of-mass energy around 250~\GeV. In this context, several focus topics have been proposed for the search of novel light scalars at such machines~\cite{Altmann:2025feg}. In this paper, we address one particular channel, i.e., the one where an additional scalar is produced that decays invisibly. Similar studies have also been presented in~\cite{deBlas:2024bmz}.

In this paper, we propose the search for additional scalars with masses $M_{S} \leq 120$~\GeV at the FCC-ee decaying invisibly. We consider a centre-of-mass energy $\sqrt{s} = 240 \rm{~GeV}$ and a luminosity of $10.8~\mathrm{ab}^{-1}$~\cite{FCC:2025lpp,FCC:2025uan, FCC:2025jtd}. Here, the new scalar is produced in association with a $Z$ boson decaying hadronically. We have also included the interference effects of the new scalar with the Standard Model Higgs boson~\cite{Robens:2026spl}, which becomes crucial for scalars with masses near the Higgs boson mass. We evaluate the expected significance and upper limit on the production of the new scalar as a function of its mass. 

After the discovery of a particle that is in agreement with the SM Higgs boson \cite{ATLAS:2022vkf,CMS:2022dwd}, phenomenological studies should be conducted taking the 125 \GeV contributions into account in all scenarios. Therefore, in this work we consider a toy model that includes contributions from the 125 \GeV scalar in addition to the new scalar we consider. We here constrain ourselves to the most minimal version that is inherited from models with a singlet extension, see e.g. \cite{Robens:2015gla,Robens:2016xkb,Ilnicka:2018def}. We then couple the dark sector to the additional real singlet in the gauge eigenbasis. SM-like couplings are therefore inherited from the doublet component, and DM-like couplings from singlet component via mixing. We consider this a more complete model with respect to studies where contributions from the 125 \GeV scalar are neglected.

A preliminary version of the work shown here was presented at the ECFA workshop and also submitted to the ECFA Study report~\cite{Altmann:2025feg}. The present work extends our previous analysis. We introduce a new toy model for simulating the signal processes. This model extends the Standard Model by introducing a scalar singlet and a dark matter candidate. Moreover, compared with our previous study, we use an improved background modelling by incorporating Initial State Radiation, Final State Radiation and also Gaussian beam spread as expected for the FCC-ee~\cite{FCC_EE_IDEA_Winter2023}.

This paper is organised as follows: in \autoref{sec:model}, we present the model that is used for studying light scalar production in this analysis; in \autoref{sec:eventgen}, we discuss the Monte Carlo samples used in the analysis and the configuration for the detector simulation. In \autoref{sec:analysis} we present the analysis, in particular discussing the selection-based approach and also a Multivariate Analysis (MVA)-based approach. In \autoref{sec:stat_analysis}, we discuss the statistical treatment considered in this paper and present the upper limit on the production of new scalars as a function of the new scalar mass.

\section{Model}\label{sec:model}

To model the signal process, that is, the production of light scalars at FCC-ee, we developed a toy model. The scalar sector of, for example, real singlet extensions with a $\mathbb{Z}_2$ symmetry, as e.g. presented in \cite{Robens:2015gla,Robens:2016xkb,Ilnicka:2018def}, can give examples of UV complete models where such scalar sectors are realised. In addition to extending the Standard Model by introducing a singlet scalar state, we also add a dark matter candidate that interacts with the additional scalar as well as the Higgs boson. The singlet scalar state is a Higgs-like state acquiring the same interactions as the Higgs boson. In this study, we consider the following potential:

\begin{equation}\label{eq:potential}
V(\Phi,S_0) = -\mu_H^2\,\Phi^\dagger\Phi + \lambda(\Phi^\dagger\Phi)^2 - \mu_S^2\,S_0^2 + \lambda_S S_0^4 + \lambda_{HS}\,\Phi^\dagger\Phi\,S_0^2
\end{equation}

Here, $\Phi$ is the SM Higgs doublet and $S_0$ is a real singlet scalar field; $\mu_H$ and $\mu_S$ are the mass parameters of the doublet and singlet sectors, $\lambda$ and $\lambda_S$ are the corresponding quartic self-couplings, and $\lambda_{HS}$ is the Higgs-singlet portal coupling.

This potential closely aligns with the model in the above literature. In particular, the number of free couplings is determined by the requirement of renormalisability. We also apply an additional $\mathbb{Z}_2$ symmetry under which all SM fields transform evenly, while the additional scalar field transforms oddly. We, however, allow for a vacuum expectation value for the latter that softly breaks the additional symmetry. This allows the scalar fields to mix.  In our scenario, both fields obtain a vacuum expectation value according to

\begin{equation}
\Phi (x)\,=\,\frac{1}{\sqrt{2}}\binom{0}{v+\varphi(x)},\,S_0(x)\,=\,\frac{1}{\sqrt{2}}\left( v_S+\varphi_S (x) \right)
\end{equation}

where we choose to stay in the unitary gauge. Here, $v$ and $v_S$ are the vacuum expectation values for the $\Phi$ and $S_0$ fields, respectively.

The mixing between the new scalar and Higgs boson is defined as:

\begin{equation}
\begin{pmatrix}
h_1 \\[1mm]
h_2
\end{pmatrix}
=
\begin{pmatrix}
\cos\alpha & \sin\alpha \\
-\sin\alpha & \cos\alpha
\end{pmatrix}
\begin{pmatrix}
\varphi_S \\[1mm]
\varphi
\end{pmatrix}
\end{equation}

where the $\sin\alpha$ parameter defines the relative strength of mixing between the scalar and Higgs boson. We identify $h_1$ as the additional scalar and $h_2$ as the Standard Model Higgs boson. For the masses satisfying $2\,M_{1} < M_{2}$, the decay process $h_2 \to h_1 h_1$ is allowed. The parameters in the potential given in \autoref{eq:potential} can be written in terms of $M_{1}$, $M_{2}$, $\sin\alpha$, $v$, $v_S$ of the model as follows~\cite{Robens:2015gla}:

\begin{equation}
\begin{aligned}
\lambda &= \frac{M_{1}^2 \cos^2\alpha + M_{2}^2 \sin^2\alpha}{2\,v^2}, \\[6pt]
\lambda_S &= \frac{M_{1}^2 \sin^2\alpha + M_{2}^2 \cos^2\alpha}{2\,v_S^2}, \\[6pt]
\lambda_{HS} &= \frac{\sin2\alpha \left(M_{2}^2 - M_{1}^2\right)}{2\,v\,v_S}, \\[6pt]
\mu_H &= \sqrt{\lambda\,v^2 + \frac{1}{2}\lambda_{HS}\,v_S^2}, \\[6pt]
\mu_S &= \sqrt{\lambda_S v_S^2 + \frac{1}{2}\lambda_{HS} v^2}.
\end{aligned}
\end{equation}

As we plan to study the decay into invisible final states, we also introduce an additional scalar particle that serves as a dark matter candidate in our model, which we label as $\chi$ in the following. The corresponding interaction term is given by \footnote{We here do not claim to discuss a full UV complete model, all discussion should be seen in a simplified approach that allows for mappings of more complete theories. The Lagrangian given here could e.g. be realised using an additional $\mathbb{Z}_2$ symmetry that remains exact.}

\begin{equation}
\mathcal{L}_{\mathrm{BSM}} = \frac{1}{2}\,\partial_\mu \chi\,\partial^\mu \chi 
- \frac{1}{2} \mu_{\chi}^2  \chi^2 
- \lambda_\chi\, S_0^2\, \chi^2
\end{equation}

After electroweak symmetry breaking, the model has in total seven free parameters in the scalar sector, which we take to be the masses of scalars ($M_{1}$ and $M_{2}$), the parameter of the dark matter candidate ($\mu_{\chi}$), the sine of the mixing angle ($\sin\alpha$), the vacuum expectation value of the doublet as well as the additional singlet ($v,\,v_S$), and finally the strength of interaction between dark matter and SM Higgs before mixing ($\lambda_\chi$):

\begin{equation}
M_{1},\ M_{2},\ M_\chi,\ \sin\alpha,\ v_S,\ v,\ \lambda_\chi.
\end{equation}

Two of these, namely $M_{2}$ and $v$, are set to the Standard Model values: $M_{2}=125$ GeV~\cite{ParticleDataGroup:2026aaa} and $v=246$ GeV. The physical mass of the dark matter candidate is given by the following relation: 

 \begin{equation}
     M_{\chi}^{2}  = \mu_{\chi}^2 + \lambda_{\chi} v_S^2
 \end{equation}

We further simplify the parameter scan by fixing the following parameters: $\sin\alpha = 0.985$, as motivated by \cite{Robens:2022zgk} 
\and $M_{\chi}=5$ \GeV. We then perform a scan over a parameter space spanning: $M_{1} \otimes v_S \otimes \lambda_{\chi}$. The ranges for the variables are as follows~\footnote{Originally the scan covered $[-4\pi,\,4\pi]$ and was refined to a smaller range after several runs.}:

\begin{equation}
\begin{aligned}
\lambda_{\chi} &\in [-10^{-1},\,-10^{-6}] \cup [10^{-6},\,10^{-1}], \quad N_{\lambda} = 400 \\
v_S &\in [50,\,5\times10^{5}]~\mathrm{GeV}, \quad N_{s} = 200\\
M_{1} &= 15 + 5n\quad [\mathrm{GeV}],\quad n \in \mathbb{Z},\quad 0 \le n \le 21.
\end{aligned}
\end{equation}

where $N_{\lambda}$ and $N_{s}$ are the number of parameter points that were considered for scanning over $\lambda_{\chi}$ and $v_S$, respectively.

To incorporate the constraint for perturbativity of coupling, we set the following criterion:

\begin{equation}
|\lambda_i| \leq 4\pi
\end{equation}

where $\lambda_i$ represents $\lambda_H,\,\lambda_S,\,\lambda_{HS}$. We also require that the branching fraction for $h_2\to \rm{invisible}$ is always less than 10\%~\cite{ATLAS:2023tkt,CMS:2023sdw}. Special attention is required for $M_{1} < 62.5$ GeV as in this case the decay channel $h_2\to h_1 h_1$ is available, allowing the decay cascade, $h_2\to h_1 h_1 \to \chi\chi\chi\chi$.  We require that $\mathcal{B}(h_2 \to \chi\chi) + \mathcal{B}(h_2\to h_1 h_1) < 10\%$, following the constraint on invisible decays of the Higgs boson from the LHC experiments. Finally, the parameter points for the analysis were selected such that the decay width of the Higgs boson is always in the range from 4 MeV to 8 MeV~\cite{ParticleDataGroup:2026aaa}. We assume the narrow width approximation, thereby requiring $\Gamma_{h_1}/M_{1} < 15\%$.

The final set of parameters selected for analysis are presented in \autoref{tab:parameter}. These parameters were obtained by fixing $\sin\alpha=0.985$, $M_{\chi} = 5$ GeV and $M_{2}=125$ \GeV.

The model is implemented in the \fr package~\cite{Christensen:2008py} and a Universal FeynRules Output (\ufo)~\cite{Darme:2023jdn} is prepared. This model is available via:\\
\begin{center}
    \url{https://github.com/amanmdesai/DM_OHSM_UFO_MODEL}
\end{center}

After placing the model directory in \MG within the \texttt{models/} directory, one can execute the following command in \MG~\cite{Alwall:2014hca} interface to import this model (\ohsm) for generating Monte Carlo samples:\\

\begin{center}
\begin{verbatim}            
                            import DM_OHSM --modelname
\end{verbatim}
\end{center}

We want to briefly comment on constraints that are typically imposed on such models. For example, this has been largely discussed for singlet scenarios in ~\cite{Robens:2015gla}. In this work, we apply a minimal approach where we do not impose all constraints that would in principle be viable for a UV complete theory. We included those which we think will limit the predictions discussed here significantly, as signal strength measurements or branching ratios to invisible final states. We keep the analysis and model investigated relatively simple such that complete studies which do have an according sector can easily use these to project onto their specific model space for a first sensitivity estimate, as e.g. done in \cite{Altmann:2025feg}.

\section{Event Generation and Detector Simulation}\label{sec:eventgen}

\begin{figure}
    \centering
    \includegraphics[width=0.55\linewidth]{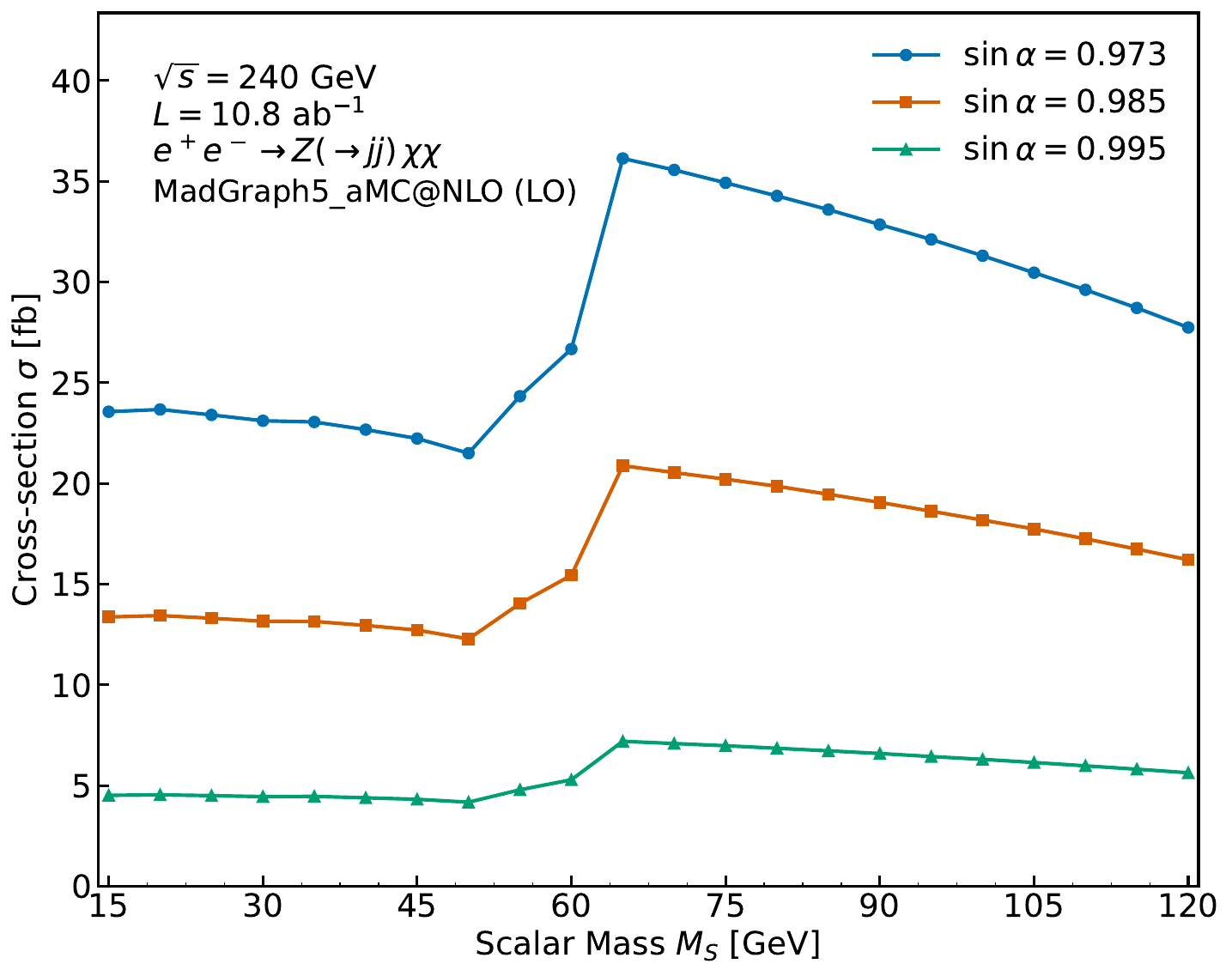}
    \caption{Production cross-section for $e^+e^- \to Z \chi\chi$ with $Z\to jj$ as a function of scalar mass at FCCee at $\sqrt{s}=240$ GeV. The cross-sections are evaluated at leading order with \MG~\cite{Alwall:2014hca} using the parameters defined in \autoref{tab:parameter}. For masses $\geq\,62.5\,\GeV$, constraints on the mixing angle are relaxed leading to larger cross sections.}
    \label{fig:xsec_mass}
\end{figure}

The signal process considered in this analysis involves the production of a new light scalar in association with a $Z$ boson~\cite{Robens:2026xfc}. The light scalar decays invisibly, while the $Z$ boson decays to a pair of quarks ($u,d,s,c,b$). The signal Monte Carlo samples were generated with \MG~\cite{Alwall:2014hca} using the \ohsm model. The events were generated considering final states, $q\bar{q}\chi\chi$, ensuring that the interference with the Higgs boson is accounted for during event generation\footnote{The importance of such effects e.g. for processes at the LHC has been studied in various papers, see e.g. \cite{Robens:2026spl} for a recent review.}. The events are generated for each mass considering the parameters defined in \autoref{tab:parameter}. The generation of hard-scattered events is followed by \pythiaN~\cite{Bierlich:2022pfr} for parton shower and hadronisation. The events are processed through the parametric response of the  \textsc{IDEA} detector~\cite{IDEAStudyGroup:2025gbt} as implemented within \textsc{Delphes}~\cite{deFavereau:2013fsa}. In the detector simulation, we treated $\chi$ as an invisible final state in addition to the SM neutrinos.  Scalar samples in the range (15, 120) GeV with an interval of 5 \GeV in their masses were simulated. For each signal sample, we generated one million events. With parameters fixed to the values as given in \autoref{tab:parameter}, we also varied $\sin\alpha=0.973,\,0.995$ to study the impact of the mixing angle on the final results. The signal processes and their cross-sections are summarised in \autoref{fig:xsec_mass}. The constraint on $\mathcal{B}(h_2\to\chi\chi) + \mathcal{B}(h_2\to h_1 h_1) < 10\%$  is relaxed when $h_2 < 2\ h_1$ since then the process $h_2\to h_1 h_1$ is not kinematically feasible. This relaxation on the constraint leads to an increase in the cross-section near $M_{S} = 62$ \GeV~ \footnote{$M_{1}$ is referred to as $M_S$ in the remainder of the paper.}. This is also reflected in the \autoref{tab:parameter} where the allowed values for $\lambda_\chi$ and $v_S$ show considerable differences in their values for $M_{S} < 60$ \GeV and $M_{S} > 60$ \GeV.

\begin{table}[htbp]
\centering
\renewcommand{\arraystretch}{1.3}
\setlength{\tabcolsep}{5pt}
\begin{tabular}{|rrrrrrrr|}
\toprule
$M_{S}$ [GeV] & $\lambda_{\chi}$ & $v_S$ [GeV] & 
$\Gamma_{h_1}$ [MeV] & $\Gamma_{h_2}$ [MeV] & 
$\mathcal{B}(h_1 \to \chi\chi)$ & 
$\mathcal{B}_{\mathrm{cascade}}$ & 
$\sigma$ [fb] \\
\midrule
15  & $-5.0 \times 10^{-4}$ & 5613.3 & 15.3 & 6.84 & 99.9 & 7.6 & 13.4 \\
20  & $-5.0 \times 10^{-4}$ & 5879.3 & 14.7 & 6.84 & 99.9 & 7.7 & 13.4 \\
25  & $-5.0 \times 10^{-4}$ & 5879.3 & 12.4 & 6.85 & 99.8 & 7.8 & 13.3 \\
30  & $-5.0 \times 10^{-4}$ & 5879.3 & 10.7 & 6.86 & 99.7 & 7.9 & 13.2 \\
35  & $-5.0 \times 10^{-4}$ & 6157.8 & 10.2 & 6.87 & 99.6 & 8.0 & 13.2 \\
40  & $-5.0 \times 10^{-4}$ & 6157.8 & 9.03 & 6.87 & 99.4 & 8.1 & 13.0 \\
45  & $-5.0 \times 10^{-4}$ & 6157.8 & 8.10 & 6.87 & 99.3 & 8.0 & 12.7 \\
50  & $-5.0 \times 10^{-4}$ & 5879.3 & 6.69 & 6.84 & 99.0 & 7.7 & 12.3 \\
55  & $-5.0 \times 10^{-4}$ & 8513.9 & 12.8 & 6.83 & 99.4 & 7.5 & 14.0 \\
60  & $-1.0 \times 10^{-3}$ & 5117.1 & 16.9 & 6.85 & 99.5 & 7.6 & 15.4 \\
65  & $-2.5 \times 10^{-3}$ & 2936.4 & 32.1 & 6.85 & 99.7 & 7.5 & 20.9 \\
70  & $-2.5 \times 10^{-3}$ & 2936.4 & 29.8 & 6.85 & 99.7 & 7.5 & 20.5 \\
75  & $-2.5 \times 10^{-3}$ & 2936.4 & 27.9 & 6.85 & 99.6 & 7.5 & 20.2 \\
80  & $-2.5 \times 10^{-3}$ & 2936.4 & 26.2 & 6.85 & 99.6 & 7.5 & 19.9 \\
85  & $-2.5 \times 10^{-3}$ & 2936.4 & 24.7 & 6.85 & 99.5 & 7.5 & 19.5 \\
90  & $-2.5 \times 10^{-3}$ & 2936.4 & 23.3 & 6.85 & 99.5 & 7.5 & 19.1 \\
95  & $-2.5 \times 10^{-3}$ & 2936.4 & 22.1 & 6.85 & 99.4 & 7.5 & 18.6 \\
100 & $-2.5 \times 10^{-3}$ & 2936.4 & 21.1 & 6.85 & 99.4 & 7.5 & 18.2 \\
105 & $-2.5 \times 10^{-3}$ & 2936.4 & 20.1 & 6.85 & 99.3 & 7.5 & 17.7 \\
110 & $-2.5 \times 10^{-3}$ & 2936.4 & 19.2 & 6.85 & 99.2 & 7.5 & 17.3 \\
115 & $-2.5 \times 10^{-3}$ & 2936.4 & 18.4 & 6.85 & 99.2 & 7.5 & 16.7 \\
120 & $-2.5 \times 10^{-3}$ & 2936.4 & 17.6 & 6.85 & 99.1 & 7.5 & 16.2 \\
\bottomrule
\end{tabular}
\caption{Benchmark points for the scalar model showing input parameters ($M_{S}$, $\lambda_\chi$, $v_S$) and resulting observables: total decay widths of $h_1$ and $h_2$, and production cross sections ($\sigma(e^+e^- \to Z \chi\chi)$) at $\sqrt{s} = 240$ GeV . For the parameter scan, we fixed $\sin\alpha=0.985$, $M_{\chi} = 5$ GeV and $M_{2} = 125$ \GeV. $\mathcal{B}_{\mathrm{cascade}}$ refers to $\mathcal{B}(h_2\to\chi\chi) + \mathcal{B}(h_2\to h_1 h_1)$. Branching ratios are given in percent.}
\label{tab:parameter}
\end{table}

The background processes used in this analysis and their cross-sections are summarised in \autoref{tab:background_xsec_nevents}. These samples were produced centrally by the FCC-ee as part of the \textsc{Winter2023} campaign~\cite{FCC_EE_IDEA_Winter2023}. The background processes arise primarily from inclusive four-fermion final states, namely $e^+e^-\to ZZ$ and $e^+e^-\to W^+W^-$, which are generated inclusively over all decay channels of the two gauge bosons, along with the $e^+e^-\to \nu\bar{\nu}Z$ process. We also consider the case where an additional lepton is generated through t-channel exchange of a charged gauge boson, with  an polar angle  $0.988 < |\cos (\theta)| < 0.999$  motivated by detector performance. To avoid double counting with the $e^+e^-\to W^+W^-$ sample, events in which the $\ell\nu$ system has an invariant mass in the window $74 < M_{\ell\nu} < 86$ GeV are excluded at generation level.  Processes as e.g. $e^+e^-\to Z/\gamma^*\to q\bar{q}$ constitute an additional  background. Further sources of background consists of Higgs-strahlung processes such as $e^+e^-\to \ell\ell H,\,q\bar{q}H,\,\nu\bar{\nu}H$, in which the Higgs boson decays invisibly or to $WW^{(*)}$/$ZZ^{(*)}$. Finally, the contribution from multiperipheral processes such as $\gamma\gamma\to q\bar{q}$~\footnote{The process single W production and $\gamma\gamma \to q\bar q$ was generated with software tools as discussed in text.}, as well as the single-photon exchange processes $e^{\pm}\gamma \to e^{\pm}Z(\to q\bar{q})$. The $e^+e^-\to Z/\gamma^*\to q\bar{q}$, $e^+e^-\to ZZ$, and $e^+e^-\to W^+W^-$ processes were generated directly using \pythiaN~\cite{Bierlich:2022pfr}. The remaining samples were all simulated by \whizard~\cite{Kilian:2007gr} and followed by \pythia~\cite{Sjostrand:2006za}. The Monte Carlo events are passed through the parametric detector simulation of the \textsc{IDEA} detector~\cite{IDEAStudyGroup:2025gbt} as implemented in \textsc{Delphes}~\cite{deFavereau:2013fsa}. The event generation and simulation are implemented within the \textsc{Key4HEP} software~\cite{Key4hep:2023nmr}. The background simulations in this analysis consider the initial state radiation, final state radiation and a Gaussian beam spread.

\begin{table}[tbp]
\centering
\renewcommand{\arraystretch}{1.3}
\begin{tabular}{|l c c|}
\toprule
\textbf{Process} & $N_{\text{events}}$ & $\sigma$ [pb] \\
\midrule
$e^+ e^- \to ZZ$ & $5.62 \times 10^{7}$ & 1.36 \\
$e^+ e^- \to W^+ W^-$ & $3.73 \times 10^{8}$ & 16.4 \\
$e^+ e^- \to Z/\gamma^* \to q\bar{q}$ & $1.01 \times 10^{8}$ & 52.6 \\
$e^+ e^- \to e^+ e^- H(\to \rm{inv.})$ & $1.20 \times 10^6$ & $7.52 \times 10^{-6}$ \\
$e^+ e^- \to \mu^+ \mu^- H(\to \rm{inv.})$ & $1.20 \times 10^6$ & $7.10 \times 10^{-6}$ \\
$e^+ e^- \to qq H(\to \rm{inv.})$ & $1.20 \times 10^6$ & $5.60 \times 10^{-5}$ \\
$e^+ e^- \to bb H(\to \rm{inv.})$ & $1.17 \times 10^6$ & $3.15 \times 10^{-5}$ \\
$e^+ e^- \to cc H(\to \rm{inv.})$ & $1.20 \times 10^6$ & $2.45 \times 10^{-5}$ \\
$e^+ e^- \to ss H(\to \rm{inv.})$ & $1.20 \times 10^6$ & $3.15 \times 10^{-5}$ \\
$e^+ e^- \to \nu\bar{\nu} H(\to ZZ)$ & $1.20 \times 10^6$ & 1.17 \\
$e^+ e^- \to \nu\bar{\nu} H(\to bb)$ & $1.20 \times 10^6$ & 26.7 \\
$e^+ e^- \to \nu\bar{\nu} H(\to cc)$ & $1.20 \times 10^6$ & 1.34 \\
$e^+ e^- \to \nu\bar{\nu} H(\to ss)$ & $1.20 \times 10^6$ & 0.011 \\
$e^+ e^- \to \nu\bar{\nu} Z$ & $2.0 \times 10^6$ & 33.2 \\
$e^+ e^- \to \nu\ell W(q\bar{q})$ & $5.0 \times 10^6$ & 0.137 \\
$e^+ e^- \to \nu\bar{\nu} H(\to WW)$ & $1.20 \times 10^6$ & 9.94 \\
$e^+ e^- \to qq H(\to WW)$ & $1.10 \times 10^6$ & 11.5 \\
$e^+ e^- \to bb H(\to WW)$ & $1.00 \times 10^6$ & 6.45 \\
$e^+ e^- \to cc H(\to WW)$ & $1.20 \times 10^6$ & 5.02 \\
$e^+ e^- \to ss H(\to WW)$ & $1.20 \times 10^6$ & 6.45 \\
$\gamma\gamma \to q\bar{q}$ & $7.9 \times 10^{6}$ & $2.44$  \\
$e^{-}\gamma \to e^{-} Z(\to q\bar{q})$ & $4.93 \times 10^{6}$ & $1.64$  \\
$\gamma e^{+} \to e^{+} Z(\to q\bar{q})$ & $5.00 \times 10^{6}$ & $1.64$  \\
\bottomrule
\end{tabular}
\caption{Background processes considered at $\sqrt{s}=240~\mathrm{GeV}$ along with the number of events and cross-sections~\cite{FCC_EE_IDEA_Winter2023}.}
\label{tab:background_xsec_nevents}
\end{table}

We adopt a right-handed coordinate system centred at the collision point of the experiment. The $y$-axis points in the upward direction, and the $x$-axis is towards the centre of the FCC-ee. The $z$-axis is along the beam axis. The azimuthal angle {\bf $\phi$} is measured from the $x$-axis, and the polar angle {\bf $\theta$} is measured from the $z$-axis. The parameterisation implemented for the IDEA detector simulation is as follows: All electrons and muons satisfying the conditions $p_T>100~\rm{MeV}$ and $|\eta| < 2.56$ are reconstructed with 100\% efficiency. Identification of electrons and muons after the stage of smearing (scattering by the detector material) is assumed to be 99\% for $E>2~\rm{GeV}$ and $|\eta| < 3$~\cite{IDEAStudyGroup:2025gbt}. 

The energy resolution assumed for the calorimeter is given by~\cite{Fabjan:2003aq,IDEAStudyGroup:2025gbt,FCC_EE_IDEA_Winter2023,ParticleDataGroup:2026aaa}:
\begin{equation}
\sigma_E = \sqrt{E^{2}\,c^{2} + E\,s^{2} + n^{2}} \, ,
\label{eq:calo_resolution}
\end{equation}
where $c = 0.005\ (0.01)$, $s = 0.03\ (0.30)~\mathrm{GeV}^{1/2}$, and $n = 0.002\ (0.05)~\mathrm{GeV}$ are the coefficients of the constant, sampling, and noise terms, respectively, for the electromagnetic (hadronic) calorimeter. Note that the numbers used here correspond to the official detector coefficients used in the \textsc{Winter2023} Campaign \cite{FCCIDEADelphesCard}. More details can be found in the above references.

The jets used in this analysis are reconstructed using the exclusive Durham $k_T$ algorithm implemented in the \textsc{FastJet} package~\cite{Ellis:1993tq,Catani:1991hj,Catani:1993hr,Cacciari:2011ma}. Here, the distance between two particles labelled $i,j$ is given by~\cite{Cacciari:2011ma}:

\begin{equation}
    d_{ij} = 2~ \mathrm{min} (E_{i}^2,E_{j}^2) (1-\cos\theta_{ij})
\end{equation}

The E-recombination scheme, in which the four-vectors of clustered particles are added, is employed, and the N-jet parameter is set to $N=2$. The resulting jets are ordered by decreasing energy. 

We note that dedicated comparison of tracker- and calorimeter-detector performance between fast and full simulation at FCC-ee based on CLD~\cite{Abidi:2025mdw} finds that the electromagnetic calorimeter resolution assumed in fast simulation can be  optimistic\footnote{We also refer the reader to references \cite{Mekala:2020zys,Thomson:2015jda} that obtain similar results for the study of the branching ratio of Higgs to invisible final states using \textsc{Delphes} and the full detector simulation, respectively, in a CLIC environment.}. 

  A full comparison including a simulation with \textsc{Geant4}~\cite{GEANT4:2002zbu} and \textsc{Delphes}~\cite{deFavereau:2013fsa} is beyond the scope of the current study.

\section{Analysis}\label{sec:analysis}

The analysis chain comprising event preselection, event reconstruction, and MVA is implemented within the \textsc{FCCAnalyses} software framework~\cite{helsens_2025_15528870}. We consider an integrated luminosity ($L$) of 10.8 ab$^{-1}$.  The preselection strategy employed in the analysis is as follows: We require that there exists a pair of jets such that its invariant mass is consistent with the $Z$ boson mass, satisfying $80 < M_{jj} < 100$ GeV.  Moreover, we require that the jets are within $|\cos\theta| < 0.9$. Additional preselection criteria on the jet clustering variables  $\sqrt{d_{23}} < 40$ GeV and $\sqrt{d_{34}} < 30$ GeV are applied to suppress events with topologies not consistent with two-jet events.  We also reject events consisting of muons or electrons with $p_{\ell} > 5$ GeV. We additionally require that $p_T^{\rm miss} > 10$ GeV. A summary of the preselection criteria is given in \autoref{tab:pre-selection}.

\begin{table}[htb!]
\centering
\renewcommand{\arraystretch}{1.3}
\begin{tabular}{|l l|}
\hline
\textbf{Category} & \textbf{Preselection Criteria} \\
\hline
Event topology & Exactly two reconstructed jets \\
$Z$ boson candidate & $80~\text{GeV} < M_Z < 100~\text{GeV}$ \\
Jet acceptance & Both jets satisfy $|\cos\theta| < 0.9$ \\
Jet clustering & $\sqrt{d_{23}} < 40~\text{GeV}$ \\
                  & $\sqrt{d_{34}} < 30~\text{GeV}$ \\
Lepton veto & No electron or muon with $p_{\ell} > 5~\text{GeV}$ \\
Missing Transverse Momentum & $p_T^{\rm miss} > 10$ GeV\\
\hline
\end{tabular}
\caption{Summary of the event preselection criteria applied in the analysis.}
\label{tab:pre-selection}
\end{table}

The reconstruction of the scalar mass is implemented following the Higgs Recoil Mass technique. In particular, the recoil mass of the new scalar is defined as: 
\begin{equation}
M_{\mathrm{Recoil}}^2 = s + m_{jj}^2 - 2 E_{jj} \sqrt{s}
\end{equation}

where $s$ is the centre-of-mass energy. Here, $m_{jj}$ and $E_{jj}$ are the invariant mass and the total energy of the pair of jets arising from the $Z$-boson decay. This method is implemented within the \textsc{FCCAnalyses} software.

\begin{figure}[t]
    \centering
    \begin{subfigure}{0.45\linewidth}
        \centering
        \includegraphics[width=\linewidth]{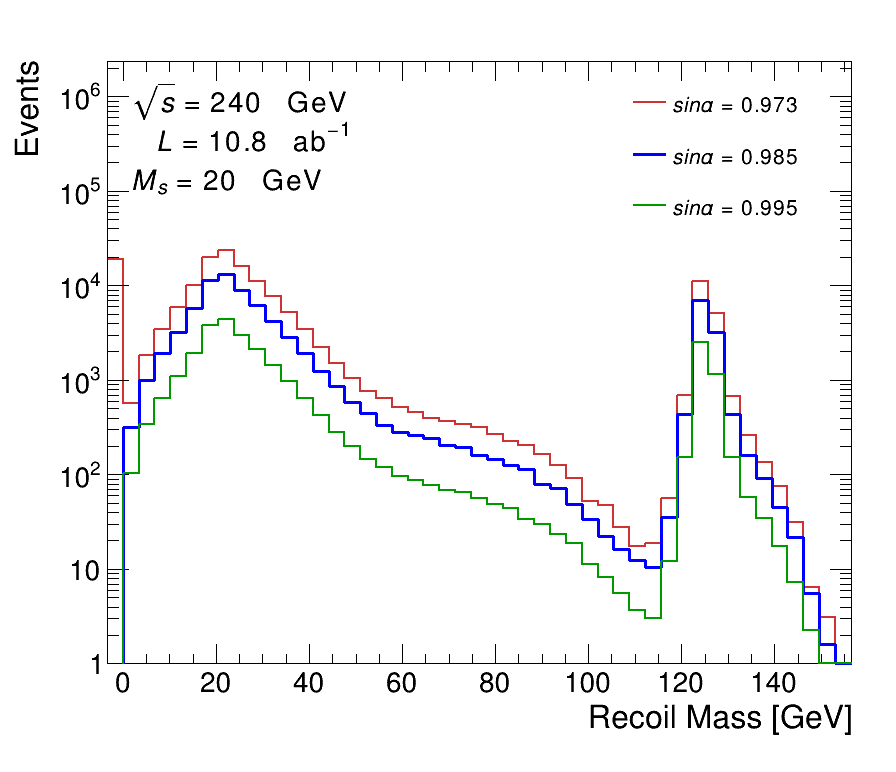}
        \caption{$M_{S} = 20$~\GeV}
    \end{subfigure}
    \hfill
    \begin{subfigure}{0.45\linewidth}
        \centering
        \includegraphics[width=\linewidth]{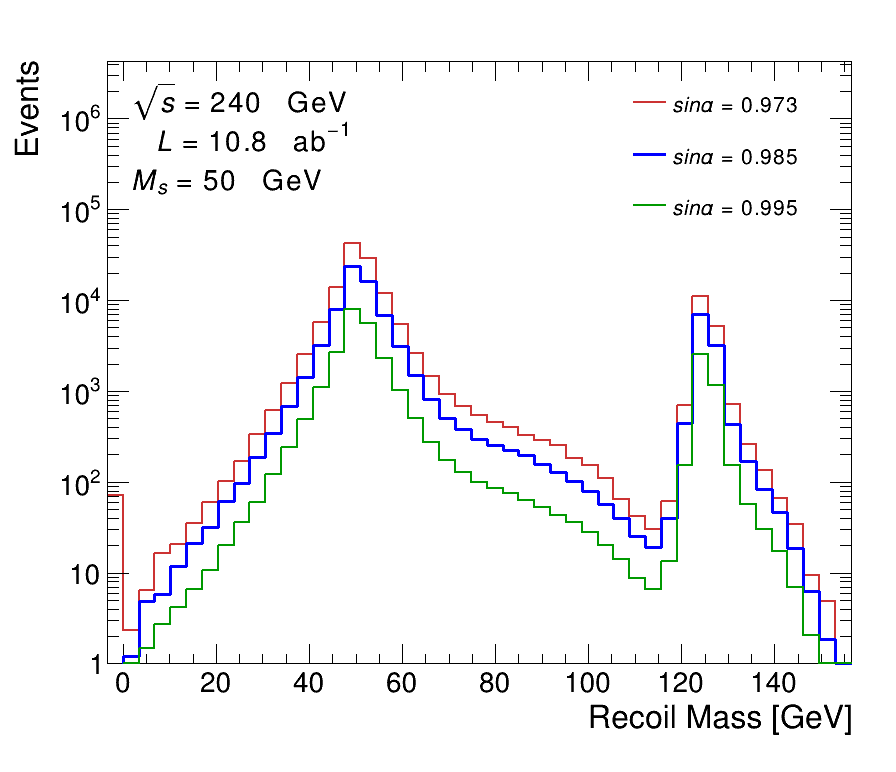}
        \caption{$M_{S} = 50$~\GeV}
    \end{subfigure}
    \hfill
    \begin{subfigure}{0.45\linewidth}
        \centering
        \includegraphics[width=\linewidth]{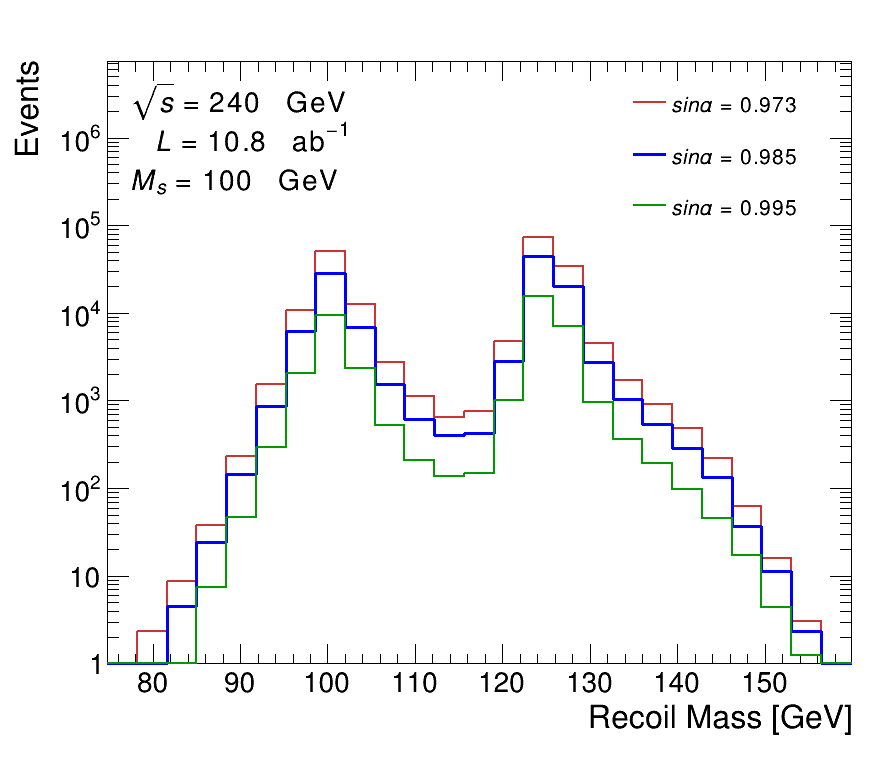}
        \caption{$M_{S} = 100$~\GeV}
    \end{subfigure}
    \hfill
    \begin{subfigure}{0.45\linewidth}
        \centering
        \includegraphics[width=\linewidth]{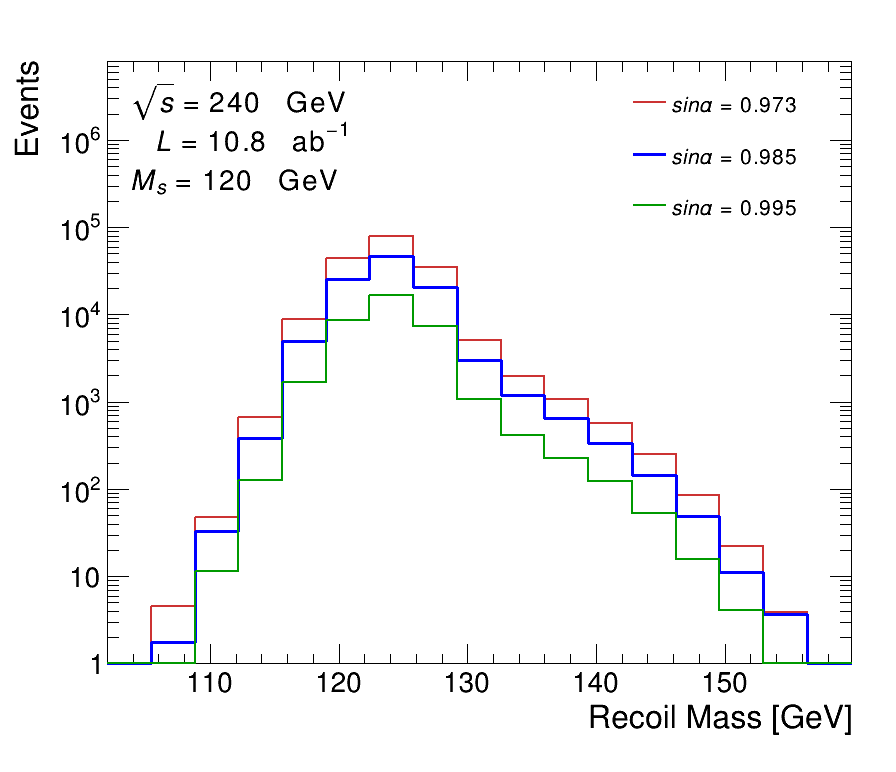}
        \caption{$M_{S} = 120$~\GeV}
    \end{subfigure}

    \vspace{0.5em}

\caption{The distribution of the recoil mass considering three different mixing angles, $\sin\alpha =0.973,\,0.985,\,0.995$. The plots are presented for $M_{S} = 20,\,50,\,100,\,120$ \GeV for the $e^+ e^- \to Z(jj)\chi\chi$ process in colours {\sl (red, blue, green)} along with the simulation for $e^+ e^- \to Z(jj)h_1(\chi\chi)$ with $\sin\alpha =0.985$ {\sl (black)}. All simulation parameters are fixed to their values given in \autoref{tab:parameter}.}
\label{fig:signal}
\end{figure}

We present the distributions of the recoil mass for four different scalar masses, $M_{S} = 20,\,50,\,100,\,120$ \GeV, in \autoref{fig:signal}. For each scalar mass point, three different mixing angles, namely, $\sin\alpha =0.973,\,0.985,\,0.995$ are considered while keeping all the other parameters fixed to the values given in \autoref{tab:parameter}. For comparison, we also display the distributions that are obtained when using $h_1$ mediation only. For lower scalar masses, the recoil mass distribution exhibits clear, distinct, well-separated peaks. As the scalar mass increases towards the Higgs boson mass, the two peaks begin to merge.  On the other hand, if one simulates the process by considering the intermediate new scalar only, then one obtains the distribution as given by the black dotted lines. In most cases these align well with the simulation in the region of the actual physical mass, but differ for higher masses. This behaviour is different for 120 \GeV where the signal alone underestimates the rate of the full contribution.

\begin{figure}[t]
    \centering

    \begin{subfigure}{0.45\linewidth}
        \centering
        \includegraphics[width=\linewidth]{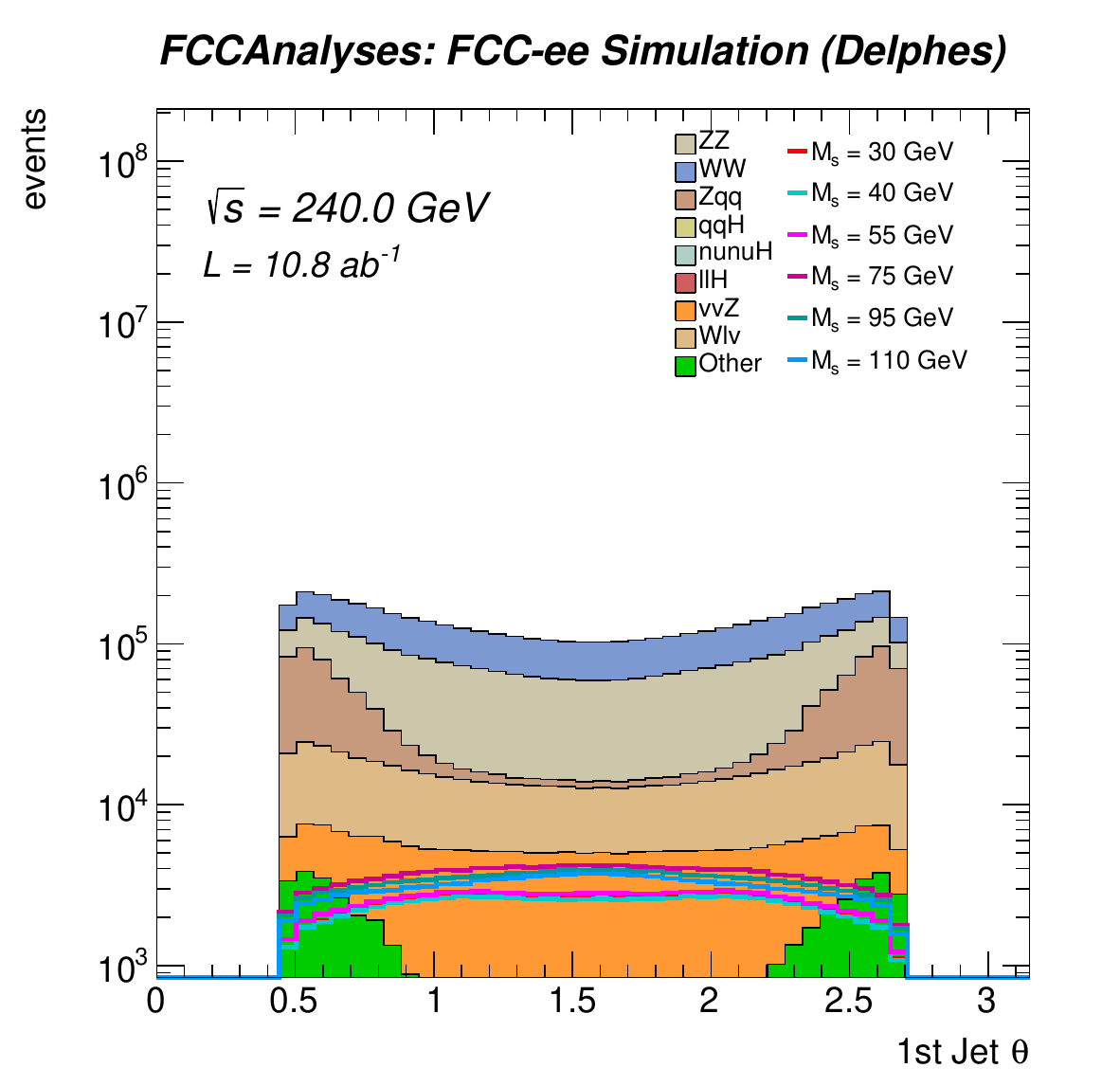}
        \caption{$\cos\theta_{j1}$}
    \end{subfigure}
    \hfill
    \begin{subfigure}{0.45\linewidth}
        \centering
        \includegraphics[width=\linewidth]{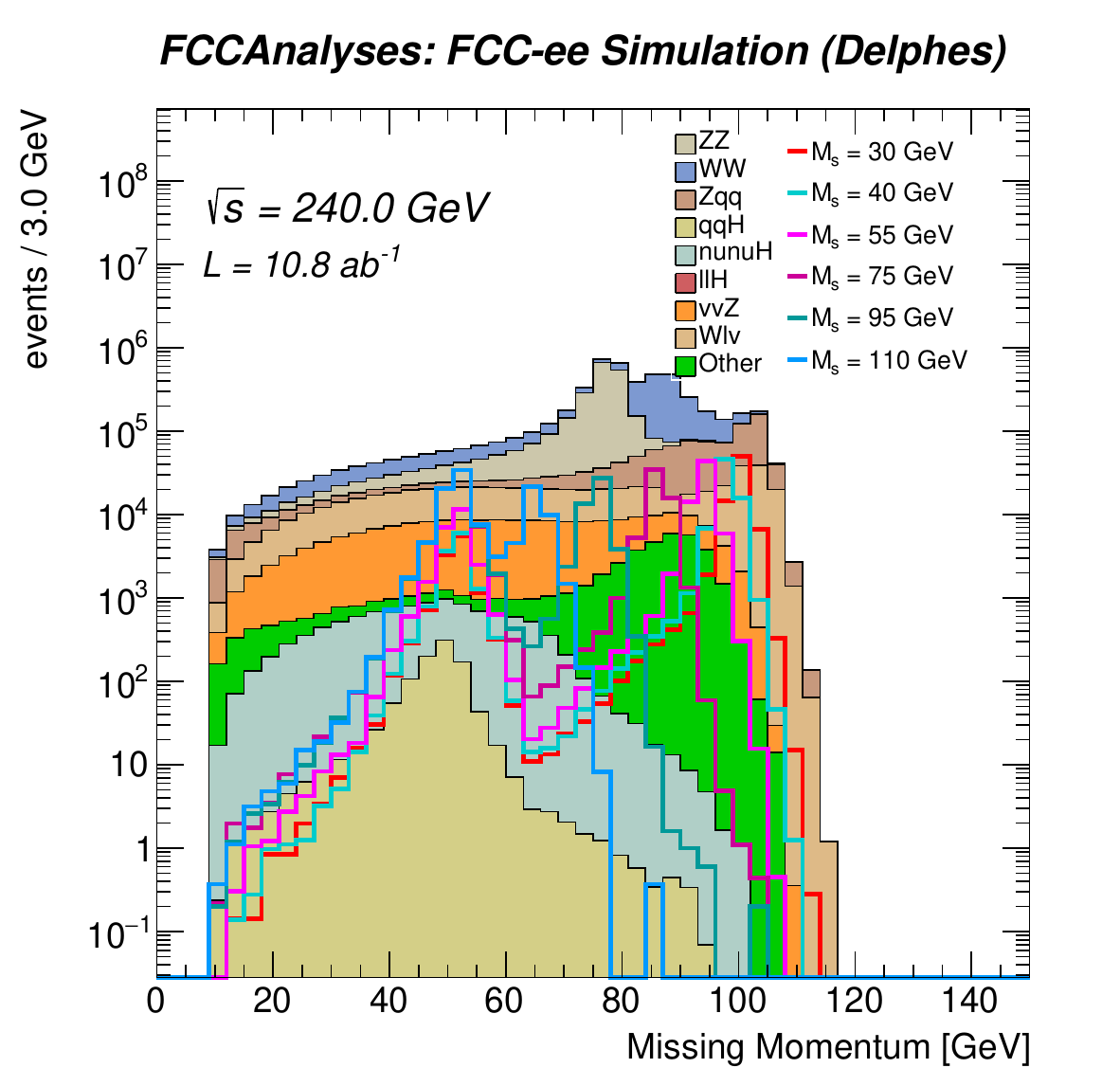}
        \caption{$p^{\mathrm{miss}}$}
    \end{subfigure}

    \vspace{0.5em}

    \begin{subfigure}{0.45\linewidth}
        \centering
        \includegraphics[width=\linewidth]{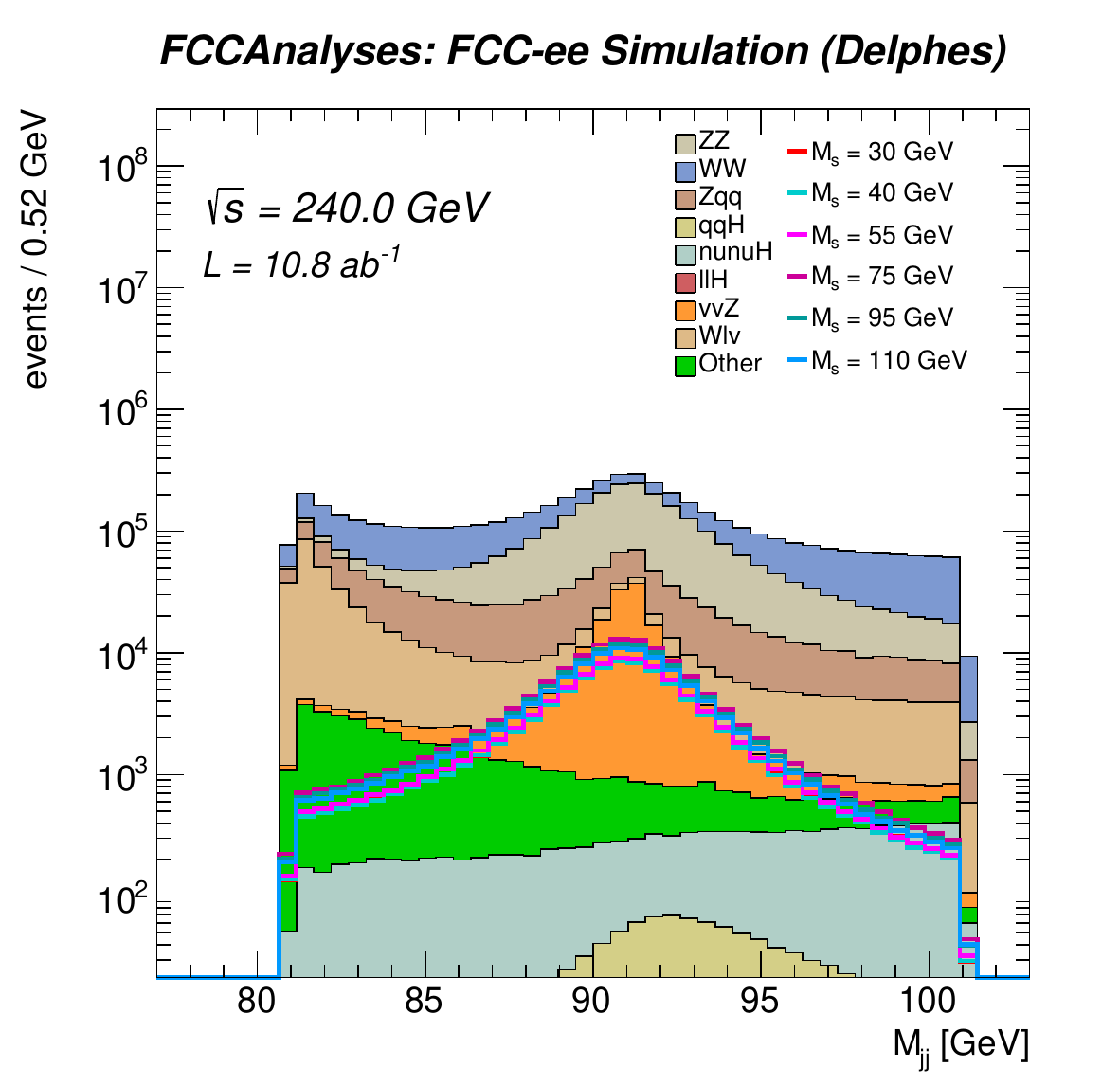}
        \caption{$M_Z$}
    \end{subfigure}
    \hfill
    \begin{subfigure}{0.45\linewidth}
        \centering
        \includegraphics[width=\linewidth]{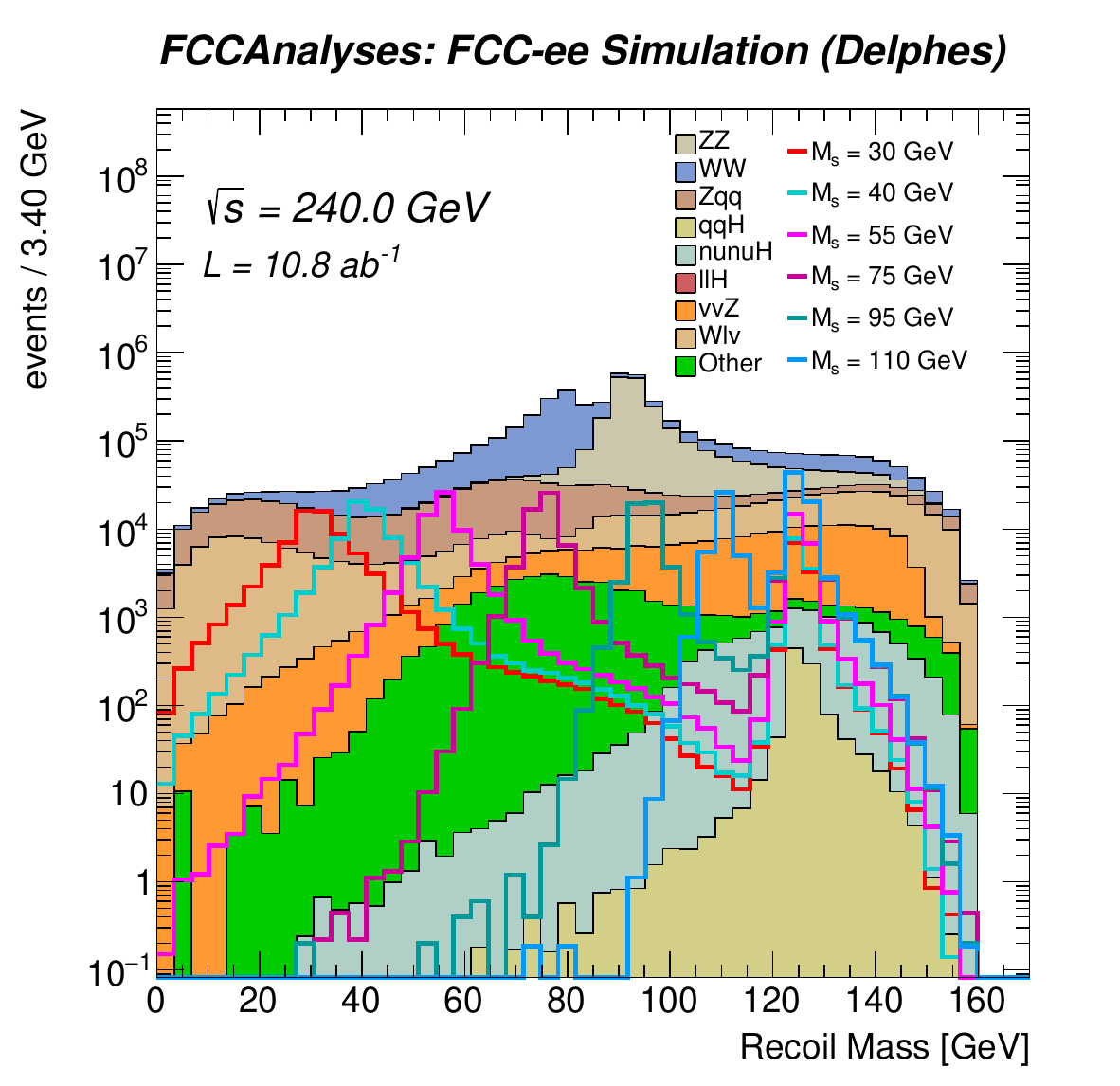}
        \caption{$M_{\mathrm{recoil}}$}
    \end{subfigure}
\caption{Distributions of (a) leading jet $\theta$, (b) missing momentum $p^{\mathrm{miss}}$, (c) dijet mass $M_{jj}$, and (d) the recoil mass $M_{\mathrm{recoil}}$ for scalar masses $M_{S} = 30, 40, 55, 75, 95,$ and $110$ \GeV, compared to the Standard Model backgrounds ($ZZ$, $WW$, $Z(q\bar{q})$, $q\bar{q}H$, $\nu\nu H$, $\ell\ell H$, and $\nu\nu Z$) after applying preselection criteria. Here $\sin\alpha =0.985$.} 
    \label{fig:preselect}
\end{figure}

The distributions of some of the variables such as the $\cos\theta_{j1}$, $p^{miss}$, $M_Z$, and $M_{\rm recoil}$ are given in \autoref{fig:preselect} for selected scalar masses. 

Each signal mass hypothesis for the process $e^+e^- \to Z \chi\chi$, is analysed separately against the full SM background. The analysis proceeds with two complementary strategies: a selection-based approach and an MVA approach. 

In the selection-based strategy, we evaluate the significance achieved for each new scalar in the study against the SM backgrounds, assuming 10\% systematic uncertainty in background modelling. Here, selections on the variables $p^{\rm miss}$ and $M_{\rm recoil}$ are considered. As the recoil mass changes for each mass and so does the missing momentum, we apply varying cuts on these variables. In particular, the missing momentum requirement is:

\begin{equation}
p^{\rm miss} >
\begin{cases}
15~\mathrm{GeV} & M_{S} = 15,\,20~\mathrm{GeV}, \\
17~\mathrm{GeV} & M_{S} = 25,\,30~\mathrm{GeV}, \\
19~\mathrm{GeV} & M_{S} = 35,\,40~\mathrm{GeV},
\end{cases}
\end{equation}
with the threshold increasing by $2~\mathrm{GeV}$ for every $10~\mathrm{GeV}$
increase in $M_{S}$.

On the other hand, we require the recoil mass to be within an interval of 15 GeV of the scalar mass.

The MVA-based analysis uses a Boosted Decision Tree (BDT) implemented in \textsc{XGBoost} software~\cite{Chen:2016btl} 
to discriminate the signal from the background. The parameters used to train the model are given in \autoref{tab:xgb_params}. We train the BDT model for each mass hypothesis against all the backgrounds listed in \autoref{tab:background_xsec_nevents}. The training is carried out over the variables summarised in \autoref{tab:ml_vars}. We use the \textsc{TMVA}~\cite{Voss:2007jxm} package within \textsc{ROOT}~\cite{Brun:1997pa} to convert the trained model into a format that can readily be applied within the \textsc{FCCAnalyses} package.

\begin{table}[tbp]
\centering
\renewcommand{\arraystretch}{1.1}
\begin{tabular}{|l c l|}
\toprule
\textbf{Parameter} & \textbf{Value} & \textbf{Description} \\
\midrule
Objective              & \texttt{binary:logistic} & Binary classification objective returning class probabilities \\
Evaluation metric      & \texttt{logloss}         & Logarithmic loss monitored during training \\
Maximum tree depth     & 5                        & Maximum depth of each decision tree \\
Learning rate          & 0.10                     & Learning rate applied during boosting \\
Subsample              & 0.8                      & Fraction of training events sampled for each tree \\
Column sample fraction & 0.8                      & Fraction of input features sampled for each tree \\
Number of estimators   & 400                      & Total number of boosted decision trees \\
Early stopping rounds  & 25                       & Training stops after 25 rounds without improvement \\
Minimum child weight   & 10                       & Minimum sum of event weights required in a leaf node \\
Gamma                  & 3                        & Minimum loss reduction required for a node split \\
Tree method            & \texttt{hist}            & Histogram-based tree construction algorithm \\
\bottomrule
\end{tabular}
\caption{\textsc{XGBoost} hyperparameters used for the binary classification.}
\label{tab:xgb_params}
\end{table}

\begin{table}
\centering
\begin{tabular}{|l l|}
\toprule
\textbf{Variable} & \textbf{Description} \\
\midrule
$E_{j_1}$ & Energy of leading jet \\
$p_{j_1}$ & Momentum magnitude of leading jet \\
$\phi_{j_1}$ & Azimuthal angle of leading jet \\
$\theta_{j_1}$ & Polar angle of leading jet \\

$E_{j_2}$ & Energy of subleading jet \\
$p_{j_2}$ & Momentum magnitude of subleading jet \\
$\phi_{j_2}$ & Azimuthal angle of subleading jet \\
$\theta_{j_2}$ & Polar angle of subleading jet \\

$p_{\text{miss}}$ & Missing momentum \\
$p_T^{\text{miss}}$ & Missing transverse momentum \\
$m_{jj}$ & Invariant mass of the dijet system \\

$E_{Z}$ & Energy of reconstructed $Z$ boson \\
$m_{Z}$ & Mass of reconstructed $Z$ boson \\
$m_{\text{recoil}}$ & Scalar recoil mass \\

$\sqrt{d_{23}}$ & Jet clustering variable \\
$\sqrt{d_{34}}$ & Jet clustering variable \\
\bottomrule
\end{tabular}
\caption{Summary of kinematic variables used in training the BDT.}
\label{tab:ml_vars}

\end{table}

\begin{figure}[htbp]
    \centering
    \includegraphics[width=0.6\linewidth]{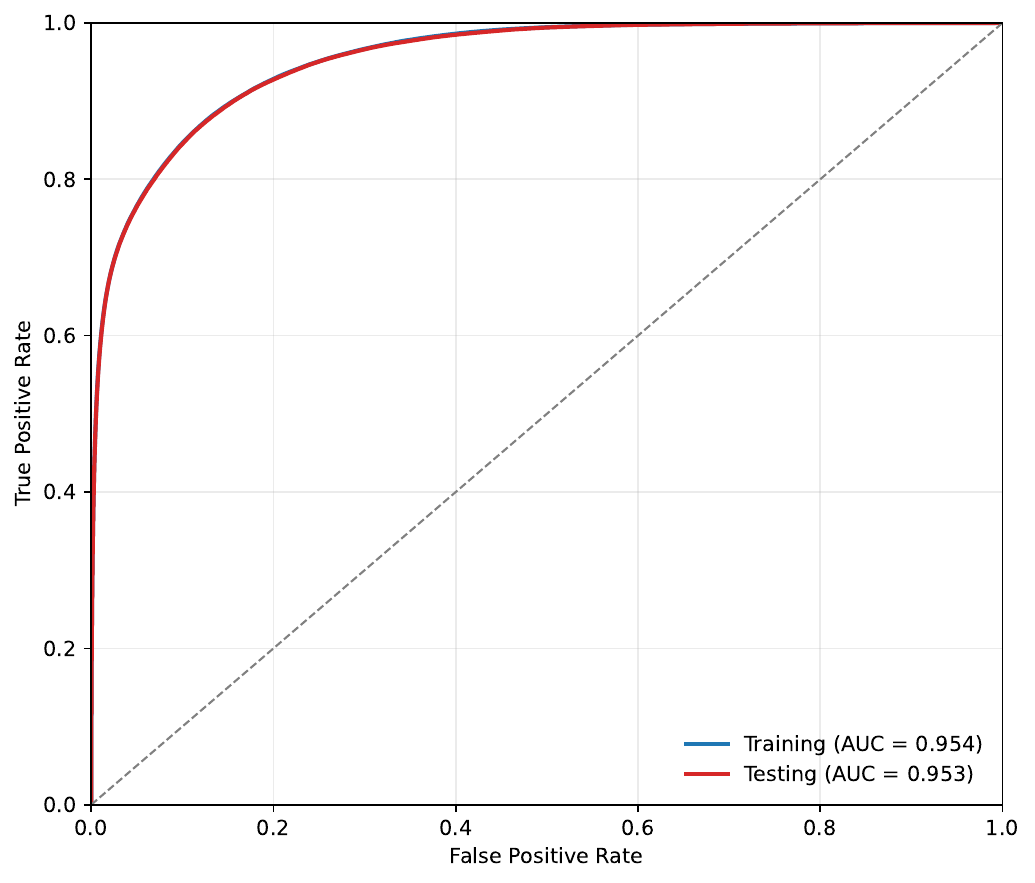}\\
    \includegraphics[width=0.75\linewidth]{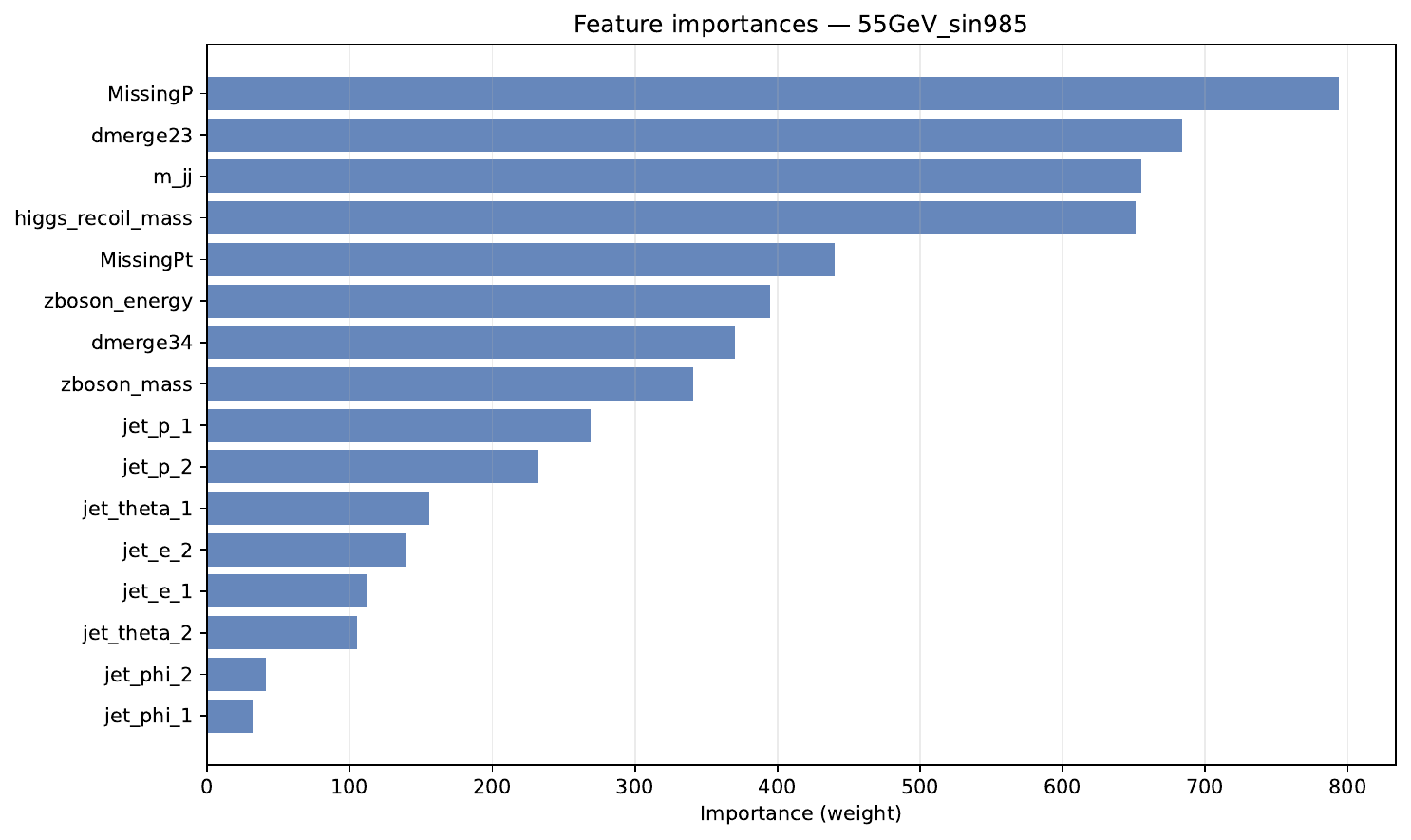}
    \caption{An example of the ROC curve for $M_{S} = 55$ GeV  with $\sin\alpha =0.985$  for both training and testing subsamples and (lower panel) the variable importance as evaluated by the BDT.}
    \label{fig:roc}
\end{figure}

Each trained model achieved an area under the receiver operating characteristic (ROC) curve greater than 0.9, with most values exceeding 0.95. This indicates that the models have a strong ability to distinguish between signal and background events. An illustrative example of a ROC curve is given in the upper panel of \autoref{fig:roc}. The ROC curve plots the false positive rate versus the true positive rate as the classifier threshold is varied, with the area under curve providing a threshold-independent measure of classification performance, where unity corresponds to perfect separation and 0.5 to random guessing. The lower panel, on the other hand, gives the relative importance of the variables which were used for training the BDT. In the case of $M_{S} = 55$ \GeV, the $p^{\rm miss}$ variable dominates in the discrimination of signal over background.

\clearpage

\begin{figure}
    \centering

    \begin{subfigure}{0.45\linewidth}
        \centering
        \includegraphics[width=\linewidth]{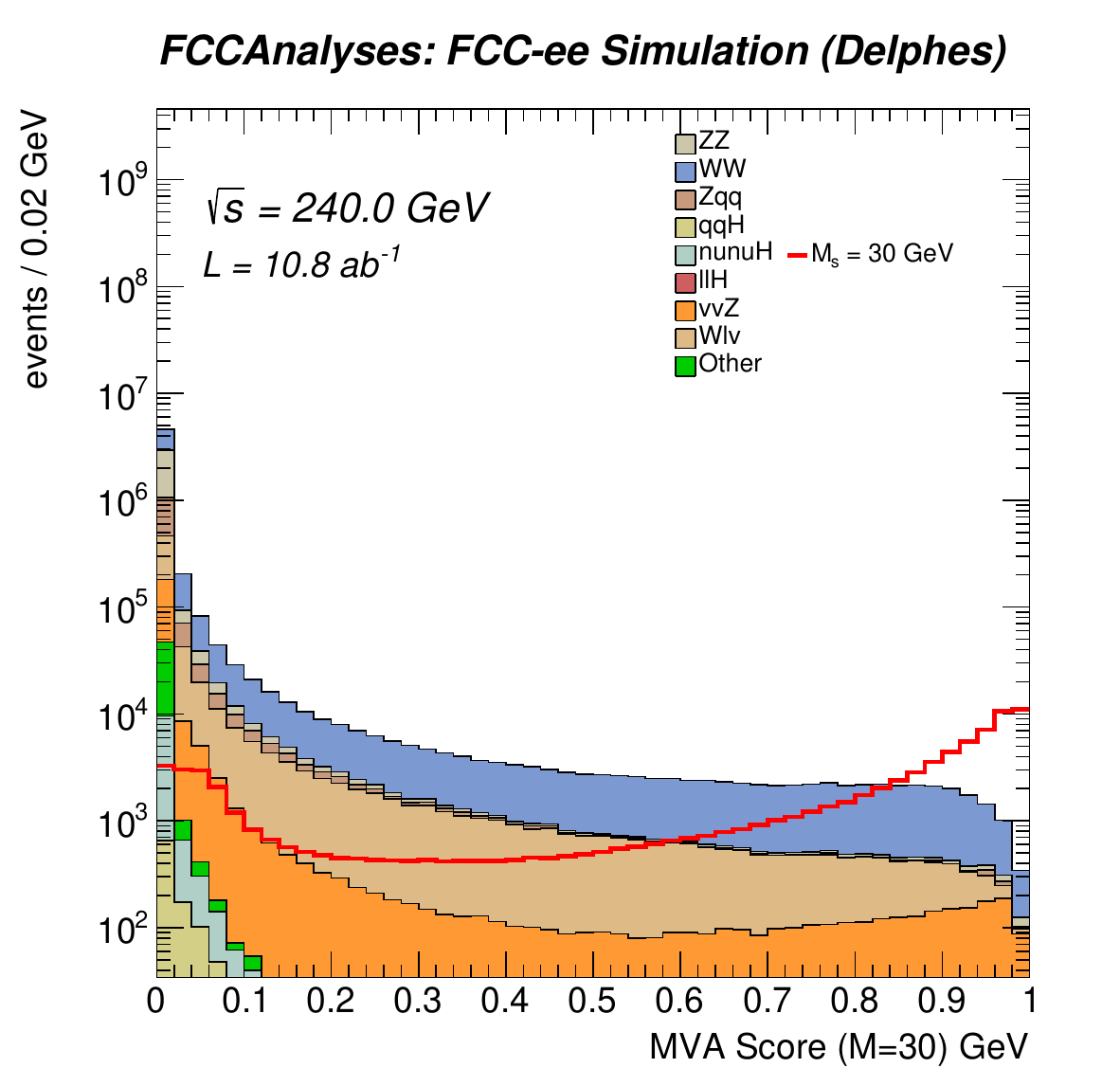}
        \caption{$M_{S} = 30$~GeV}
    \end{subfigure}
    \hfill
    \begin{subfigure}{0.45\linewidth}
        \centering
        \includegraphics[width=\linewidth]{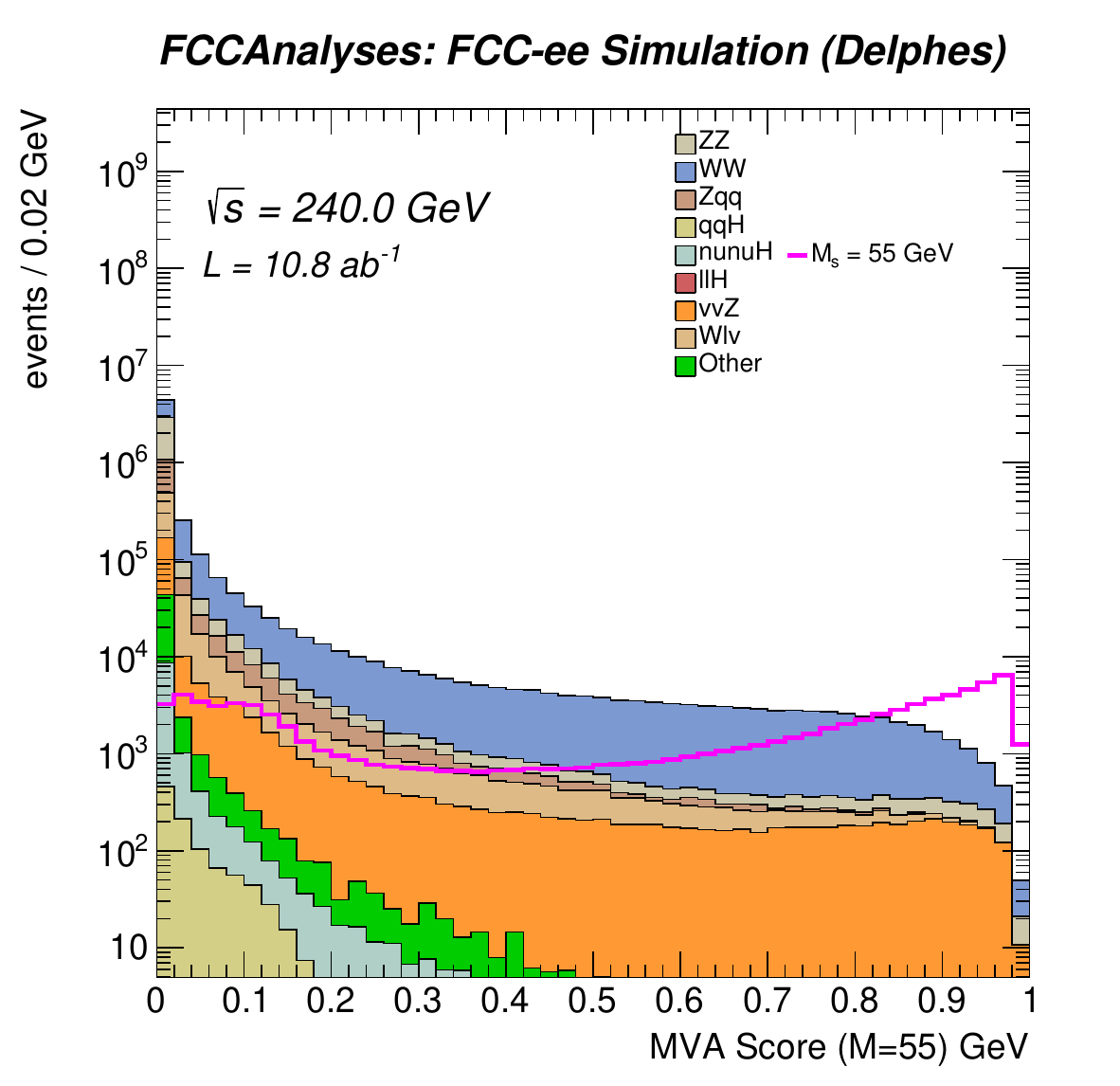}
        \caption{$M_{S} = 55$~GeV}
    \end{subfigure}

    \vspace{0.5cm}

    \begin{subfigure}{0.45\linewidth}
        \centering
        \includegraphics[width=\linewidth]{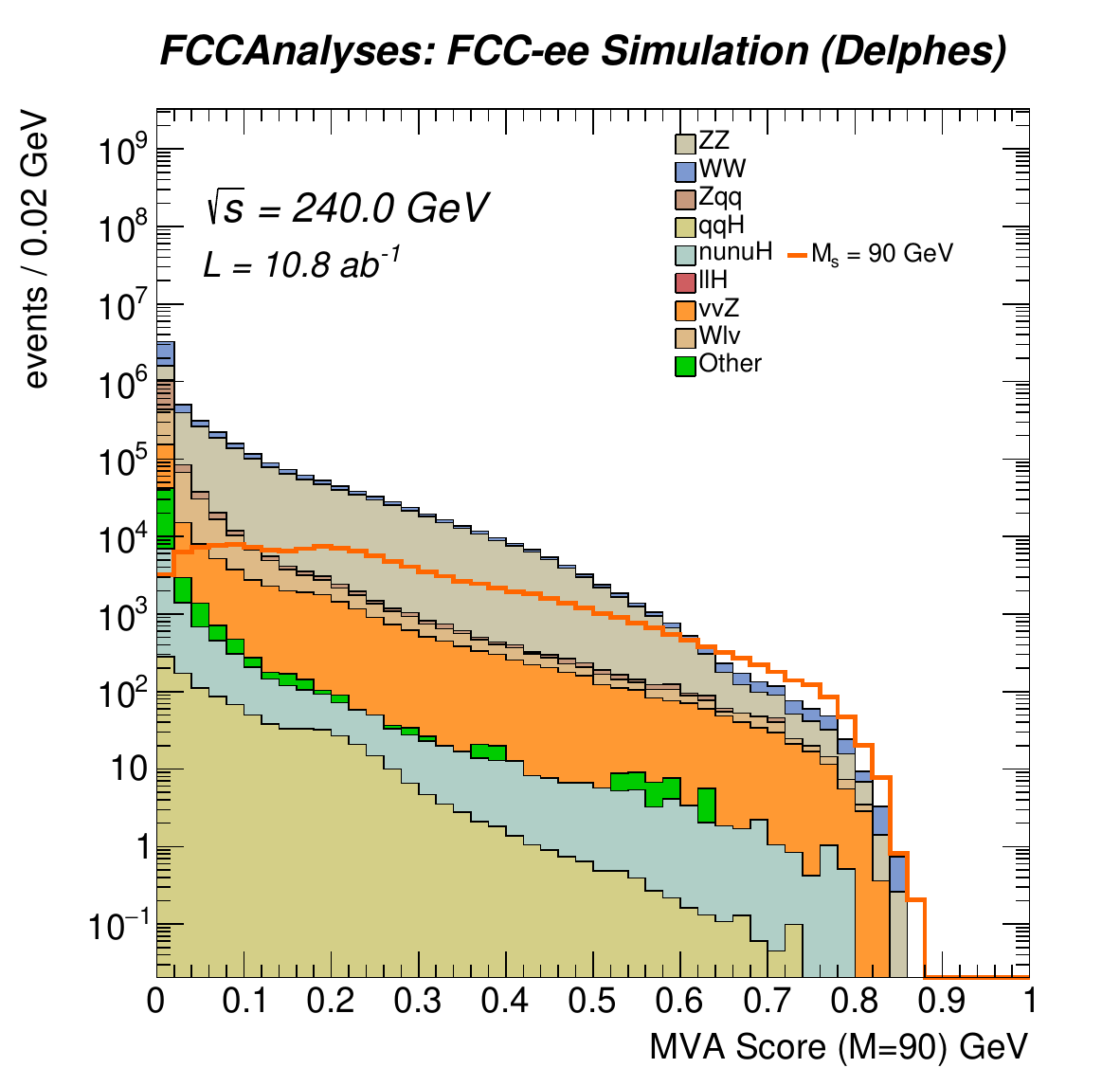}
        \caption{$M_{S} = 90$~GeV}
    \end{subfigure}
    \hfill
    \begin{subfigure}{0.45\linewidth}
        \centering
        \includegraphics[width=\linewidth]{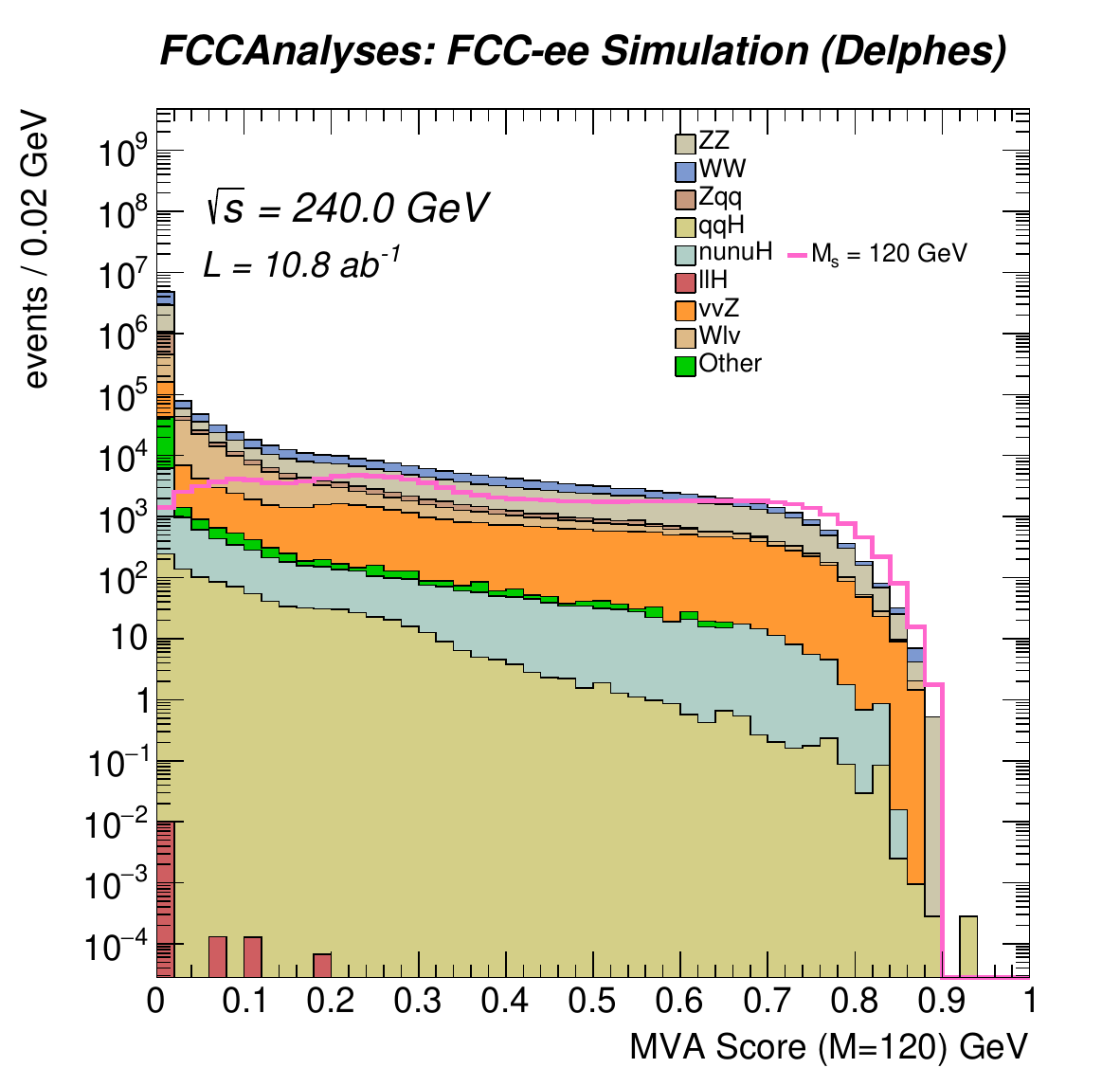}
        \caption{$M_{S} = 120$~GeV}
    \end{subfigure}

    \caption{MVA score distributions for signal samples at different scalar masses  with $\sin\alpha =0.985$, illustrating the classifier response across mass hypotheses. Subfigures correspond to (a) $30$~GeV, (b) $55$~GeV, (c) $90$~GeV, and (d) $120$~GeV.}
    \label{fig:bdt_output}
\end{figure}

The MVA score for the samples is obtained after inferring the trained model on the samples. The distribution of this variable is summarised for some selected mass hypotheses in \autoref{fig:bdt_output}. As seen in \autoref{fig:bdt_output}, the low-mass scalars ($M_{S} < 80$ GeV) are well separated from the background. However, scalars with mass near the $Z$ boson mass are affected by the background present in MVA bins with score higher than 0.8.

\begin{figure}
    \centering
    \includegraphics[width=0.7\linewidth]{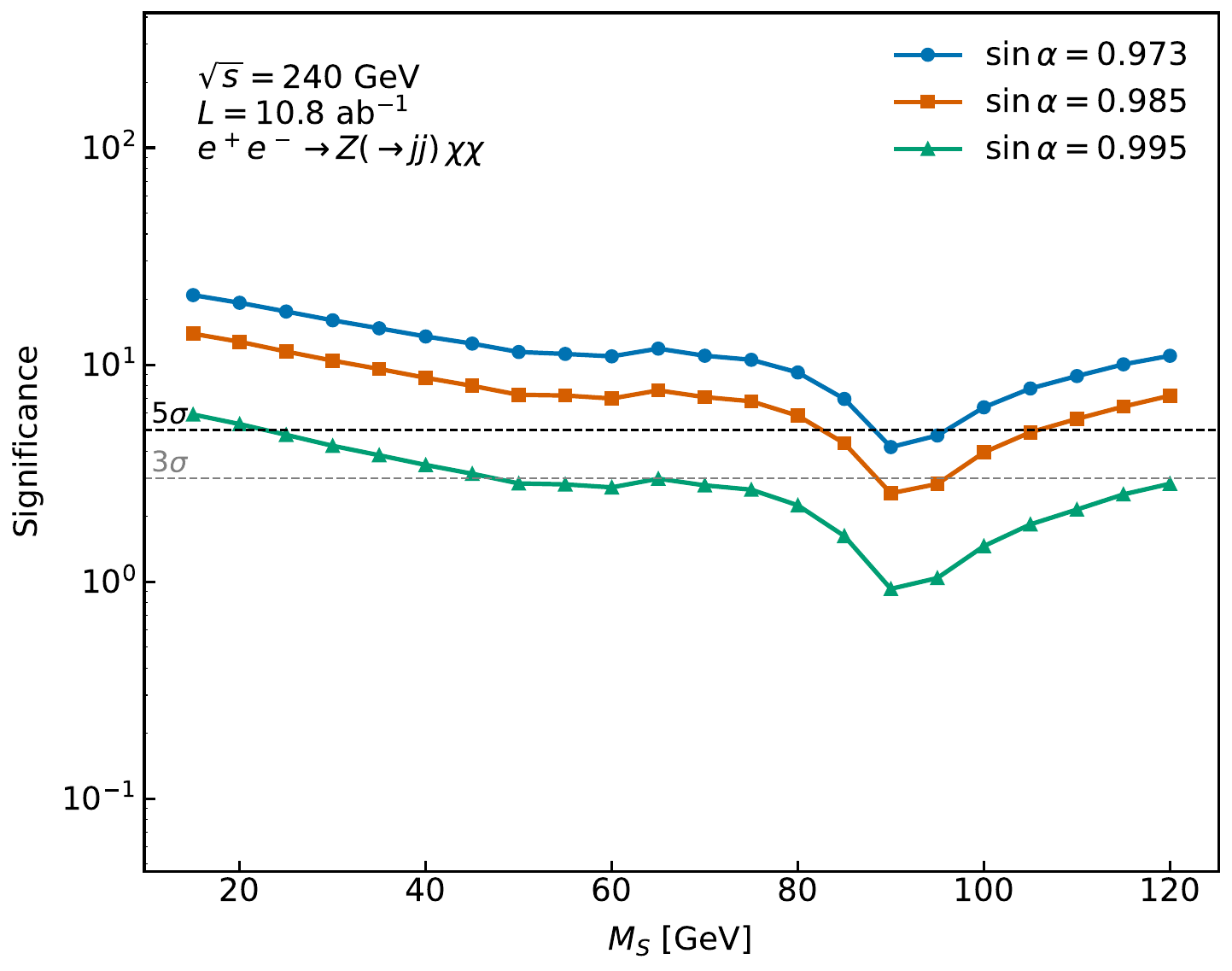}
    \caption{Expected signal significance (Asimov) for the $e^+ e^- \to Z(jj)h_1(\chi\chi)$ as a function of the scalar mass $M_{S}$. Also shown are lines for significance equal to 3 ({\sl grey}) and 5 ({\sl black}), respectively.} 
    \label{fig:significance}
\end{figure}

Additionally, we complement the selection-based strategy with the MVA-based strategy by incorporating a final selection on the MVA score. We require the events to satisfy $\rm{MVA}>0.5$. We use the Asimov significance ($Z_A$) to quantify the expected sensitivity of the signal over background, taking into account the relative background uncertainty~\cite{Cowan:2010js}:

\begin{equation}
Z _A= \sqrt{ 2 \Bigg[ 
(S + B) \ln \frac{(S + B)(B + \sigma_B^2)}{B^2 + (S + B)\sigma_B^2} 
- \frac{B^2}{\sigma_B^2} \ln \Big( 1 + \frac{\sigma_B^2 S}{B (B + \sigma_B^2)} \Big) 
\Bigg] }, \quad
\end{equation}

where $S$, $B$ represent the signal and background yields, and $\sigma_B=\delta_{\rm unc}\times B$, where $\delta_{\rm unc}$ is the uncertainty (\%) in background modelling. In this analysis, we assume a $\delta_{\rm unc}=10\%$ uncertainty in background modelling. The expected Asimov significance for the scalar signal versus SM background as a function of scalar masses is given in \autoref{fig:significance}. The signal is normalised to the cross-section and branching fraction evaluated using the parameters given in \autoref{tab:parameter}. We find that for lower scalar masses, the significance for observation is higher owing to the absence of large backgrounds. A dip in the significance is seen near the $Z$ mass owing to the high number of expected events coming from $ZZ\to q\bar{q}\nu\bar{\nu}$. The significance recovers for masses beyond the $Z$-pole, up to 120 \GeV. In general, for mixing angles $\leq\,0.99$ novel scalars can be discovered with masses up to 80 \GeV, as well as in the mass range 110--120 \GeV, according to our study.

\section{Statistical Analysis}\label{sec:stat_analysis}

We compute the expected upper limit at 95\% CL on $\sigma (e^+ e^- \rightarrow q\bar{q} h_1)\times \mathcal{B}(h_1\rightarrow \text{inv.})$. To compute this quantity, we use the \textsc{CMS Combine} framework~\cite{CMS:2024onh}, which was developed by the CMS collaboration for statistical computation.  

The limits are obtained by using the \texttt{AsymptoticLimits} function implemented in the \textsc{CMS Combine} software. This software provides asymptotic approximations to the profile likelihood ratio test statistic~\cite{Cowan:2010js} to compute limits. The evaluations are all carried out using the MVA distribution obtained by applying the trained BDT to the samples. On this variable, we only apply the selections on the missing momentum and the recoil mass. Here, the backgrounds are normalised to the SM values as given in \autoref{tab:background_xsec_nevents}, while the signal is normalised to a reference value. We assume a normalisation uncertainty of 10\% on background and additionally a 1\% uncertainty on luminosity. Systematic uncertainties were incorporated into the likelihood as log-normal nuisance parameters. The statistical analysis further considers the Monte Carlo statistical uncertainties arising due to the finite size samples~\cite{Barlow:1993dm}. We then compute the upper limit at 95\% CL on $\sigma(e^+e^- \rightarrow q\bar q h_1)\,\mathcal{B}(h_1\rightarrow \mathrm{inv})$.

\begin{figure}
        \centering
        \includegraphics[width=0.65\linewidth]{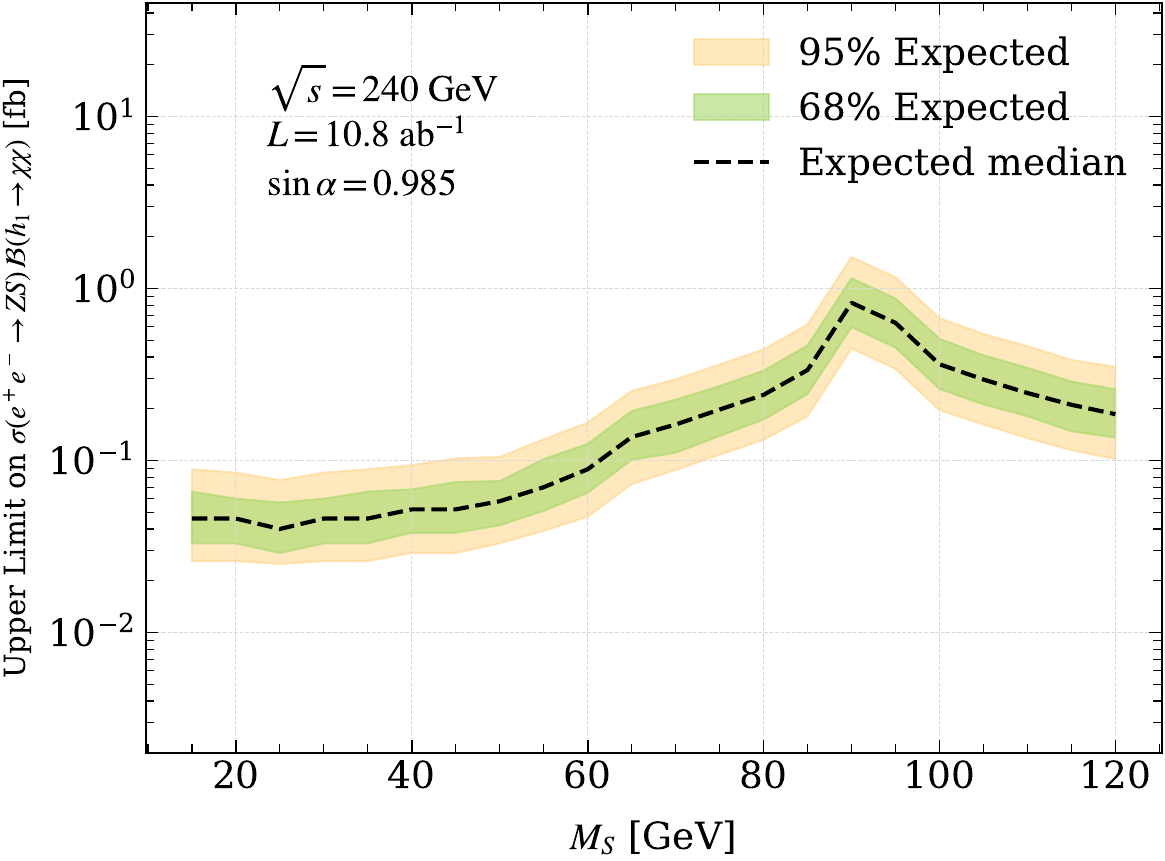}
    \caption{Expected 95\% CL upper limits on $\sigma (e^+ e^- \rightarrow q\bar{q} h_1)\times \mathcal{B}(h_1\rightarrow \text{inv.})$ for $\sin\alpha=0.985$ along with the $\pm 1\sigma$ and $\pm 2\sigma$ intervals as a function of the scalar mass.}
    \label{fig:bdt_limit_bands}
\end{figure}

We report the expected limits in \autoref{fig:bdt_limit_bands}, where we have also included the $\pm 1\sigma$ and $\pm 2\sigma$ bands in addition to the median expected limits.  We note that higher sensitivity is achieved for masses below $\sim 60$ GeV. However, for the region around the $Z$ mass, one finds the sensitivity is reduced. This feature is due to the background from $ZZ$. A further loss of sensitivity is observed for scalar masses approaching the SM Higgs boson mass, where the interference between the new scalar and the 125~\GeV Higgs boson, included in the signal generation (\autoref{sec:eventgen}), suppresses the effective signal cross-section and correspondingly weakens the resulting limits.

\begin{figure}
        \centering
        \includegraphics[width=0.65\linewidth]{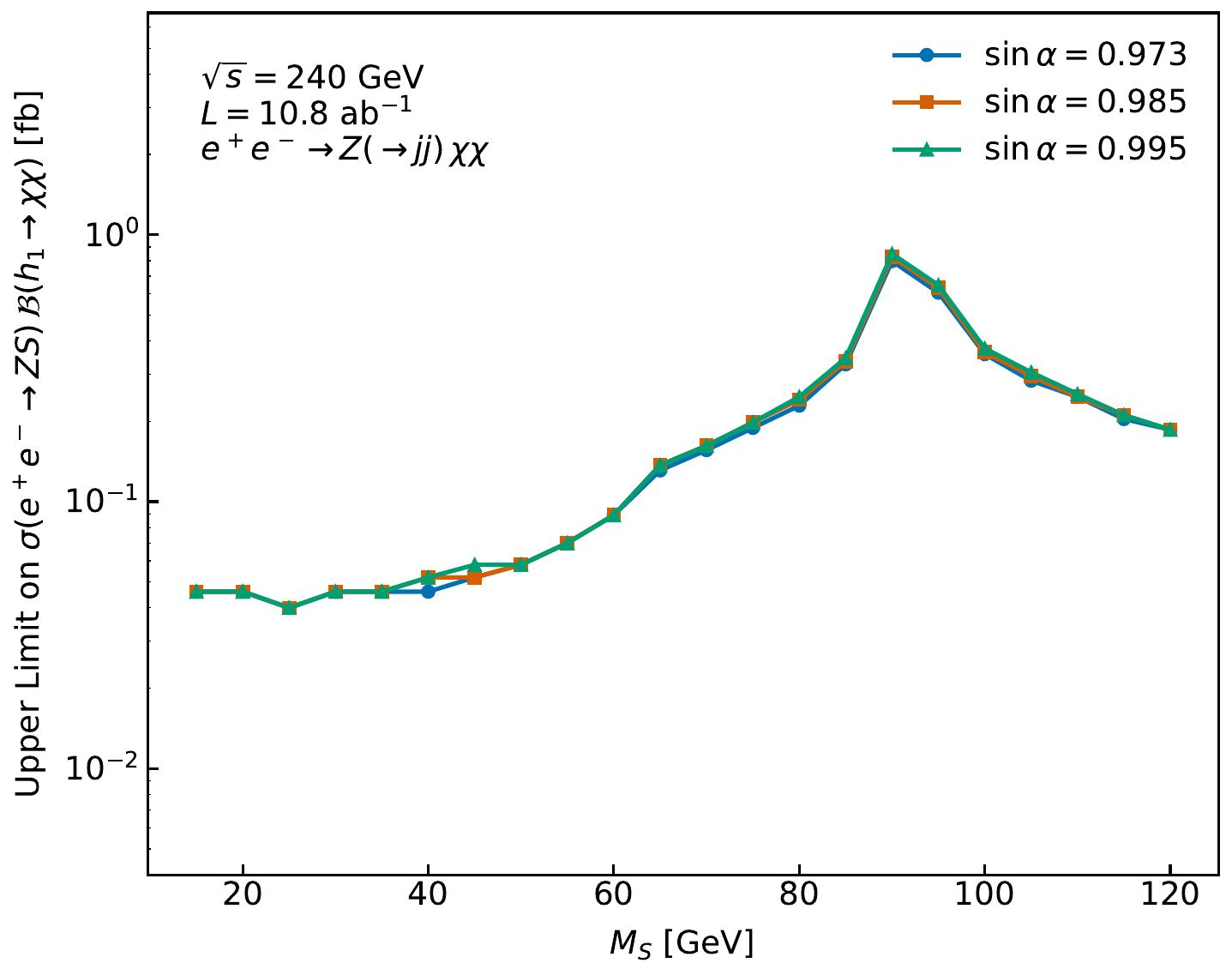}
    \caption{Expected 95\% CL upper limits on $\sigma (e^+ e^- \rightarrow q\bar{q} h_1)\times \mathcal{B}(h_1\rightarrow \text{inv.})$  as a function of the scalar mass for three mixing angle parameters, namely $\sin\alpha=0.973,\,0.985,\,0.995$.}
    \label{fig:bdt_limit}
\end{figure}

In \autoref{fig:bdt_limit}, we present the limits obtained by varying the mixing angle ($\sin\alpha$). Through this computation, we find that the limits derived in this analysis are largely independent of the mixing angle.

\begin{figure}
    \centering
    \includegraphics[width=0.65\linewidth]{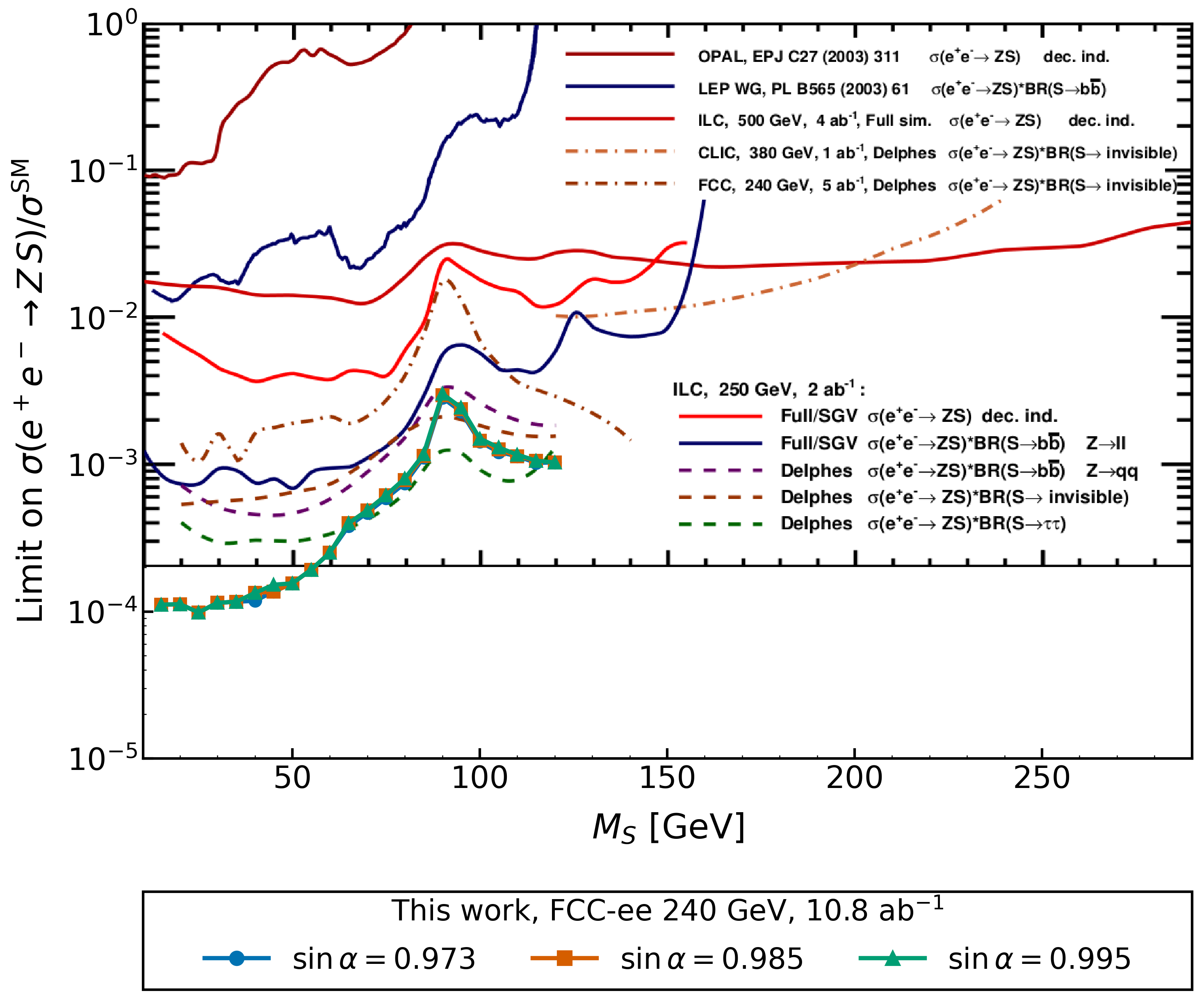}
    \caption{Expected 95\% CL upper limits on $\sigma(e^+e^-\to ZS)/\sigma^{\rm SM}$ from this work (FCC-ee, 240~\GeV, $10.8$~\ab), for $\sin\alpha = 0.973, 0.985, 0.995$, overlaid on ECFA report Figure~104 of Ref.~\cite{Altmann:2025feg}.}
    \label{fig:ecfa_overlay}
\end{figure}

In \autoref{fig:ecfa_overlay}, we further compare our results with the ILC projections at $\sqrt{s}=250$~\GeV with $2$~\ab reported in Figure 104 of Ref.~\cite{Altmann:2025feg}. Below the $Z$ pole, the expected upper limit obtained in this work is better by a factor of $2$--$3$. Near and above the $Z$ pole, owing to the interference with the SM Higgs boson, the limits are less sensitive\footnote{As many details of that and our study differ, a detailed comparison would be needed to better understand the actual differences. This is beyond the scope of the current work.}.

We remark that the previous study for FCC-ee at an integrated luminosity of 5~\ab was a preliminary version of the current analysis. For the work presented here, several analysis features, in particular with respect to the usage of XGBoost and signal selection optimization, were improved.

\section{Conclusion}

We studied the production of a new light scalar in association with a $Z$ boson at the Future Circular Collider-ee (FCC-ee) at $\sqrt{s} = 240$ GeV. The $e^+ e^- \rightarrow Z \chi\chi$ process was studied for a set of scalar masses in the range $(15, 120)$ GeV. 

In contrast to many other studies, we do not neglect possible contributions from the 125 \GeV scalar state. We consider this a more realistic approach as most scenarios with novel scalars which exhibit mixing between the new scalars and the SM-like scalar when going from the gauge to the mass eigenstates. Our toy model therefore includes all possible contributions in a scenario where the existence of the 125 \GeV is not neglected.

A parametric simulation of the IDEA detector was used to incorporate detector effects.  In addition to using a selection-based strategy, we also applied a Multivariate Analysis technique, in particular using a BDT from \textsc{XGBoost} to distinguish signal from background. We evaluated the expected Asimov significance as a function of scalar mass, finding higher sensitivity for masses well below the $Z$ boson mass, while reduced sensitivity is observed near the $Z$ boson mass due to the dominant $ZZ$ background, with the significance recovering for masses beyond the $Z$-pole up to 120 \GeV. We also find that, depending on the mixing angle, novel scalars with masses below 80 \GeV, as well as with masses in the range 110--120 \GeV, are within the discovery reach. We derived expected upper limits on the production cross-section times branching fraction at 95\% CL, using the MVA score distribution after applying selections on the missing momentum and the recoil mass. Sensitivity up to $\sim 10^{-2}$--$10^{-1}$~fb could be achieved for scalars with mass below $\sim 60$ GeV, while sensitivities in the range $0.1$--$1$~fb could be achieved for masses between $\sim 60$ and $120$ GeV, with the weakest sensitivity near the $Z$-pole at $M_{S}\approx 90$ GeV. The derived limits are found to be largely independent of the scalar--Higgs mixing angle.

Finally, our expected limits are, below the $Z$ pole, about a factor of $2$--$3$ better than the ILC projections of Ref.~\cite{Altmann:2025feg}, with our limits moving closer to the ILC projections near and above the $Z$ pole where the interference with the Higgs boson is sizeable as also seen from \autoref{fig:signal}.

\section*{Acknowledgements}
\noindent We thank Prof. Aleksander Filip Zarnecki for useful discussions. TR acknowledges financial support from the Croatian Science Foundation (HRZZ) project ``Beyond the Standard Model discovery and Standard Model precision at LHC
Run III”, IP-2022-10-2520.

\bibliographystyle{JHEP}
\bibliography{biblio}

\end{document}